# AN ENERGY-EFFICIENT BENNETT CLOCKING SCHEME FOR 4-STATE MULTIFERROIC LOGIC

Noel D'Souza[1], Jayasimha Atulasimha[1], Supriyo Bandyopadhyay[2]

1) Department of Mechanical and Nuclear Engineering,

2) Department of Electrical and Computer Engineering,

Virginia Commonwealth University, Richmond, VA 23284, USA.

Email: {dsouzanm, jatulasimha, sbandy}@vcu.edu

Nanomagnets with biaxial magnetocrystalline anisotropy have four stable magnetization orientations that can encode 4-state logic bits (00), (01), (11) and (10). Recently, a 4-state NOR gate derived from three such nanomagnets, interacting via dipole interaction, was proposed. Here, we devise a Bennett clocking scheme to propagate 4-state logic bits unidirectionally between such gates. The nanomagnets are assumed to be made of 2-phase strain-coupled magnetostrictive/piezoelectric multiferroic elements, such as nickel and lead zirconate titanate (PZT). A small voltage of 200 mV applied across the piezoelectric layer can generate enough mechanical stress in the magnetostrictive layer to rotate its magnetization away from one of the four stable orientations and implement Bennett clocking. We show that a particular sequence of positive and negative voltages will propagate 4-state logic bits unidirectionally down a chain of such multiferroic nanomagnets for logic flow.

## 1. INTRODUCTION

An emerging technology in the field of digital computing is nanomagnetic logic (NML) that promises significantly lower power dissipation than conventional transistor-based electronics [1, 2, 3, 4, 5, 6]. NML is non-volatile, which permits implementation in both logic and memory, and it has no standby power dissipation unlike transistors.

In conventional *binary* NML, bits 0 and 1 are encoded in two stable magnetization directions of single-domain nanomagnets with uniaxial shape anisotropy [1, 2]. Data transmission between them requires: (i) dipole interaction between neighbors, and (ii) a Bennett clock that temporarily reorients the magnetization of every nanomagnet away from one of the stable directions to allow a bit to propagate through it [3, 7, 8]. The re-orientation can be carried out with an effective magnetic field that is generated either with an external current that does not pass through the nanomagnet but produces a local magnetic field in its vicinity [6, 9], or with a spin-polarized current that passes through the nanomagnets and generates a spin transfer torque [10] (or perhaps domain wall motion [11, 12]), or with a voltage that produces mechanical strain in a magnetostrictive-piezoelectric multiferroic nanomagnet [13, 14, 15, 16] and rotates its magnetization vector. The latter switching modality, termed "straintronics", promises unprecedented energy efficiency and is the subject of this paper. We show how straintronics can be employed to "Bennett clock" unconventional *multi-state* (specifically 4-state) logic circuits in NML with extremely low energy dissipation.

One way to implement Bennett clocking in traditional binary NML is to arrange shape anisotropic nanomagnets in a line along their hard axis as shown in Fig. 1(a). The ground state of the array will be "anti-ferromagnetic" whereby each nanomagnet's magnetization will align along the easy axis, but nearest neighbors will have anti-parallel magnetizations, representing a sequence of binary bits 0 1 0 1…. This anti-ferromagnetic ordering happens because of dipole interaction between neighbors. If we now flip the first nanomagnet's magnetization (first bit) with some external agent and expect all succeeding nanomagnets to sequentially flip in a domino-fashion to re-assume the anti-ferromagnetic order because of dipole interaction, that will not happen. What prevents its occurrence is that immediately after switching the first nanomagnet, the second nanomagnet finds itself in a frustrated state where its left neighbor's dipole interaction and right neighbor's dipole interaction exactly cancel. Therefore, this nanomagnet does not flip and the input bit does not propagate further.

In order to break this logjam and make the input bit propagate, one needs a clock [17] to manipulate the dipole interactions between neighboring pairs of nanomagnets. For example, prior to flipping the first bit, a global magnetic field could break the anti-ferromagnetic ordering and align every nanomagnet's magnetization along the common hard axis. This field is then withdrawn and the magnetization of the first nanomagnet is oriented by an external agent to conform to the input bit [18]. Dipole interaction will then flip the magnetization of all the succeeding nanomagnets sequentially in a domino-like fashion since every nanomagnet now experiences non-zero dipole interaction that restores the anti-ferromagnetic order. This is an example of propagating bits using Bennett clocking. Here the global magnetic field acts as the clock. The same type of clock can propagate an input bit down a chain if the nanomagnets are arranged in a line parallel to the easy axis as shown in Fig. 1(b). In this case, dipole coupling will cause ferromagnetic ordering.

Unfortunately, the use of a global magnetic field makes the architecture non-pipelined and hence unacceptably slow and error-prone [2]. A superior strategy is to employ local clocking where the orientation of every nanomagnet is turned along the hard axis with a local agent one at a time to implement Bennett clocking [3]. This increases the lithography overhead significantly since every nanomagnet needs to be contacted, but it allows pipelining of data and makes the architecture much faster [2]. Since non-pipelined and error prone architectures are unacceptable, we will consider only the local clocking scheme. The issue then is what constitutes a suitable agent for local clocking, i.e. what is the most energy efficient way to rotate the magnetization of a nanomagnet from the easy to the hard axis? Obviously, this can be achieved with a local magnetic field directed along the hard axis that is generated with a current passing through a nearby wire [6, 9], or with spin transfer torque (STT) current passing through the nanomagnet [10, 12], or with a small voltage generating a stress in a multiferroic nanomagnet [13, 14, 15, 16, 19, 20, 21, 22, 23].

The three possible methods for local clocking differ vastly in their energy-efficiencies. The first consumes an exorbitant amount of energy and is most energy inefficient. Ref. [9] employed this scheme and dissipated at least $10^{12}$ kT ($4.2 \times 10^{-9}$ Joules) per bit flip while operating at a clock rate of ~ 1 MHz at room temperature. This is an experimental result, but even theoretically, the energy dissipation is not likely to be any less than $10^8$ kT per bit flip for this methodology. For the second method (STT), ref. [12] reported a dissipation of at least $10^4$ kT at a clock rate of 500 MHz. Once again, this is an experimental result, but the theoretical estimate is not significantly better. Finally, the third method ("straintronics") may turn out to be most energy-efficient. Although no experimental result is available (except for demonstrating the basic mechanism of switching the magnetization of a multiferroic with voltage-generated strain [19]), there are theoretical estimates that claim energy dissipation of only ~ 200 kT to switch a multiferroic nanomagnet at clock rates exceeding 1 GHz [13, 14, 16]. Here, we study straintronic clocking schemes to propagate composite logic bits (2-bit states) in 4-state logic circuits.

## 2. THEORY

A typical multiferroic nanomagnet that can encode 4-state logic bits is illustrated in Fig. 2, with the magnetostrictive layer (nickel) on top and the piezoelectric layer (lead zirconate titanate, or PZT) at the bottom. The shape is circular, so that there is no shape anisotropy. The biaxial magnetocrystalline anisotropy in the magnetostrictive layer then creates four possible stable magnetization directions ("up", "right", "down", "left") in which 2-bit states (00, 01, 11, 10) are encoded, as shown in Fig. 2. This encoding scheme results in a change of only a single bit for every 90° magnetization rotation.

Because of magnetocrystalline anisotropy, unstressed single-crystal Ni has its "easy" axis of magnetization along the $<111>$ direction, a "medium" axis along the $<110>$ direction and a "hard" axis along the $<100>$ direction. In our study, we assume that the two-dimensional geometry of the Ni layer suppresses out-of-plane excursion of the magnetization vector because of the large magnetostatic energy penalty, so that the magnetization vector always lies in the (001) plane. In that case, the "easy" axes of single-crystal Ni in the unstressed state are [110], [$\bar{1}\bar{1}0$], [$\bar{1}10$] and [$1\bar{1}0$] in Miller notation, as illustrated in the saddle-shaped curve in Fig. 2 [15]. The coordinate axis system was rotated by 45° in order to have the stable states/bits point "up", "right", "down", "left" along the x- and y-axes. Thus, the [100] axis lies along the +45° direction.

This paper studies a synchronous Bennett clocking scheme where each 4-state multiferroic nanomagnet is subjected to a particular stress cycle that will allow 4-state logic bits to be propagated unidirectionally along a data path. We develop a novel scheme for such logic propagation and demonstrate its feasibility by modeling the rotation of magnetization of each nanomagnet due to a cycle of tensile and compressive stresses generated by positive and negative electrostatic potentials applied across the piezoelectric layer of each multiferroic nanomagnet.

## 3. RESULTS AND DISCUSSION

When magnetizations of two adjacent nanomagnets are parallel to the line joining their centers, the ordering will be ferromagnetic, but when the magnetizations are perpendicular to this line, the ordering will be anti-ferromagnetic because of dipole interaction. Thus, if the first bit in a linear array of circular 4-state multiferroics is switched from its initial state to one of the three other stable states, three possible arrangements result. Since each nanomagnet has four possible magnetization orientations, there are twelve distinct configurations that may arise when the first bit is switched, as illustrated in Fig. 3.

Consider the anti-ferromagnetic arrangement of Fig. 3(a), with the first nanomagnet's magnetization orientation acting as the input bit to the line. In this configuration, the input magnetization can be switched from its initial "up" state to the "down", "right" or "left" state. The corresponding nanomagnet states are shown in the **Final State** column (Fig. 3(a)) based on the fact that coupling will be ferromagnetic (F) along the nanomagnet-array axis and anti-ferromagnetic (AF) perpendicular to this axis. Therefore, when the input bit is flipped from "up" to "down", the change is propagated along the array if it is appropriately clocked, with the input magnetization direction replicated in every odd-numbered nanomagnet from the left. This is a consequence of anti-ferromagnetic ordering. If the input is switched to either "left" or "right", ferromagnetic coupling will ensure that all the nanomagnets assume the "left" or "right" orientation, respectively. Similar considerations apply to the other three configurations in Figs. 3(b) – 3(d). Here, we only present the numerical results corresponding to row I in the arrangements of Figs. 3(a) which pertains to the anti-ferromagnetic arrangement with the input magnetization oriented "up". All other cases have been exhaustively examined to confirm successful operation, but are not presented here due to space limitations.

Fig. 3(a) shows that in an anti-ferromagnetically coupled line, the first bit will be replicated in every odd-numbered nanomagnet (and has therefore propagated through the line) if the array *can reach ground state* after the first bit is flipped. This can happen only if the array does not get stuck in a metastable state and fail to reach the ground state [17, 24]. It can be shown that dipole interaction alone cannot guarantee that the ground state will be reached, which is why multi-phase clocking is needed to nudge the system out of any metastable state should the system get stuck in one [17]. Additionally, the dipole interaction energy is usually not sufficient to overcome the magnetocrystalline anisotropy energy and rotate a nanomagnet out of its current orientation to a different orientation in order to propagate the bit. Thus, once again, a clock is needed to supply the energy needed to overcome the magnetocrystalline anisotropy energy. In Bennett clocking schemes, the clocking agent (local magnetic field, spin transfer torque, strain, etc.) will rotate the magnetization into an unstable state, perching it at the top of the energy barrier, and then the dipole interaction of its neighbors will push it into the desired stable state, thus ensuring unidirectional propagation of a logic bit. All this can happen reliably if we neglect thermal fluctuations that can induce errors in switching. The effect of thermal fluctuations is beyond the scope of this paper, but preliminary considerations show that they will undoubtedly induce errors at room temperature, but not to the point where the scheme is invalidated.

Consider the nanomagnet array of Fig. 4 consisting of four nanomagnets in the collective ground state of the array (row I). The magnetization of nanomagnet 1 on the far left is the input bit. If it is flipped from its initial "up" to "down" state at time $t = 0$, then at time $t = 0+$, we reach the situation shown in row II where nanomagnet 2 experiences equal and opposite dipole interactions from its two nearest neighbors (magnets 1 and 3) which are magnetized in opposite directions. As a result, the net dipole interaction experienced by nanomagnet 2 is zero. Thus, this nanomagnet does not flip its magnetization in response to the first nanomagnet's flip, preventing propagation of the input logic bit down the chain. In other words, the array is stuck in a *metastable state* and cannot reach the ground state.

In order to break this logjam and allow the logic bit to flow past nanomagnet 2, we have to apply the following clock cycle. We will assume that nanomagnets 1 and 4 remain stiff while nanomagnets 2 and 3 rotate when stressed. This is a good approximation if the magnetocrystalline anisotropy energy is significantly larger than the dipole interaction energy.

Stage 1: Tension (T)/Compression (C) (row III): After the input nanomagnet has been switched, nanomagnet 2 is subject to a tensile stress (gradually increased to a maximum value of +100 MPa), applied along the [100] direction (+45° to the +x-axis) (row III). Since Ni has negative magnetostriction, a tensile stress tends to raise the energy along the axis of applied stress while lowering the energy along the axis perpendicular to this direction. A compressive stress does the exact opposite [27]. As a result, tension applied on nanomagnet 2 along the [100] direction will prefer to align the magnetization along either -45° or +135° (-225°) directions while raising the energy barrier in the +45° and -135° (+225°) directions. Since the initial state of nanomagnet 2 is along the -90° direction and the energy barrier is raised along the -135° (+225°) direction, the only possible magnetization rotation that can take place is from -90° to -45°. Energy profiles showing the raising and lowering of energy levels of nanomagnets 2 and 3 are presented later.

At the same time, a compressive stress (gradually increased to a maximum value of -100 MPa) is applied on nanomagnet 3 along the [100] axis, which causes its magnetization to rotate from the initial +90° state to the +45° state (row III). In all cases studied in this paper, stresses are simultaneously applied on nanomagnets 2 and 3.

Stage 2: Relaxation(R)/Compression(C) (row IV): Next, the tensile stress on nanomagnet 2 is gradually reduced to zero while keeping the compressive stress on nanomagnet 3 fixed. The magnetization of nanomagnet 3 remains oriented in the +45° direction, but the magnetization of nanomagnet 2 rotates from -45° to 0°. This can be understood from the energy profiles of the nanomagnets under stress, which we discuss later. Rotations take place to lower the energy of a nanomagnet to the minimum energy state.

Stage 3: Compression(C)/Tension (T) (row V): A compressive stress (up to -100 MPa) is now applied on nanomagnet 2 and simultaneously the compressive stress on nanomagnet 3 is relaxed to zero. This is immediately followed by the application of a tensile stress (up to +100 MPa) on nanomagnet 3 while keeping nanomagnet 2 unstressed. Nanomagnet 2 rotates to its preferred lowest-energy state along +45°. The relaxation of stress on nanomagnet 3 pushes its magnetization towards 0° (ferromagnetic coupling is preferred over anti-ferromagnetic coupling since the former has a stronger dipole interaction) while the subsequent tensile stress results in rotation of the magnetization to -45°.

Stage 4: Relaxation(R)/Tension(T) (row VI): Finally, the compressive stress on nanomagnet 2 is relaxed while keeping the tensile stress on nanomagnet 3 fixed. This results in the magnetization of nanomagnet 2 rotating to the final desired state of +90° ("up").

The above clocking sequence successfully flips the magnetization of nanomagnet 2 in response to the flipping of the input nanomagnet 1 and allows the logic bit to propagate past nanomagnet 2. The same sequence of stresses is then applied to the next set of nanomagnets (3 and 4, with 2 and 5 now assumed to be stiff), which results in nanomagnet 3 eventually settling in the "down" orientation (-90°), mirroring the state of the input bit. By continuing this cycle, the input bit can be propagated down the entire chain, resulting in successful logic propagation.

In order to prove rigorously that the magnetizations of the stressed multiferroic nanomagnets orient as described, a theoretical analysis is performed to determine the energy profiles of nanomagnets 2 and 3 under stress. The total energy of any nanomagnet is given by the equation

$$E_{total} = E_{dipole} + E_{magntocrystalline} + E_{stress-anisotropy}, ,  \qquad (1)$$

where $E_{dipole}$ is the dipole-interaction energy due to neighboring nanomagnets, $E_{magnetocrystalline}$ is the intrinsic magnetocrystalline anisotropy energy and $E_{stress-anisotropy}$ is the stress anisotropy energy introduced by stress applied along the [100] direction. Since the shape of the nanomagnet is isotropic, there is no shape anisotropy energy.

After nanomagnet 1 is switched, and nanomagnets 2 and 3 are stressed, their magnetizations rotate in order to reach the minimum energy state. Let us assume that their magnetization vectors subtend angles $\theta_2$ and $\theta_3$ with the x-axis. In order to find these angles for the minimum energy state under a given stress, we make two simplifying assumptions: First, we assume that the magnetocrystalline anisotropy energy is so much larger than the dipole interaction energy that nanomagnets 1 and 4 are immune to dipole influences of their neighbors and do not rotate when nanomagnets 2 and 3 rotate under stress. Second, we will assume that the stresses on the nanomagnets are changed slowly enough that their magnetization vectors can follow quasi-statically. In that case, it is sufficient to compute the energy minima of nanomagnets 2 and 3 ($E_{total-2}$ and $E_{total-3}$) under any arbitrary stress to find the angles $\theta_2$ and $\theta_3$. There is also a third assumption here; namely, that we neglect effects of thermal fluctuations that may drive the system out of its minimum energy state randomly.

The total energies of nanomagnets 2 and 3 are given by

$$E_{total-2} = \frac{\mu_0 M_s^2 \Omega^2}{4\pi R^3}[-2\cos\theta_2(\cos\theta_3 + \cos\theta_1) + \sin\theta_2(\sin\theta_3 + \sin\theta_1)]$$

$$+ \frac{K_1\Omega}{4}\cos^2(2\theta_2) - \frac{3}{2}\lambda_{100}\sigma\Omega\cos^2\left(\theta_2 - \frac{\pi}{4}\right)$$

$$E_{total-3} = \frac{\mu_0 M_s^2 \Omega^2}{4\pi R^3}[-2\cos\theta_3(\cos\theta_4 + \cos\theta_2) + \sin\theta_3(\sin\theta_4 + \sin\theta_2)]$$

$$+ \frac{K_1\Omega}{4}\cos^2(2\theta_3) - \frac{3}{2}\lambda_{100}\sigma\Omega\cos^2\left(\theta_3 - \frac{\pi}{4}\right), \quad (2)$$

where the first term is the dipole-interaction energy of a nanomagnet with its neighbors, the second term is the magnetocrystalline anisotropic energy, and the third term is stress anisotropy energy resulting from a stress $\sigma$ applied along the [100] direction (45° with the x-axis). Here, $M_s$ is the saturation magnetization, $\Omega$ is the nanomagnet's volume, $\mu_0$ is the permeability of free space, $R$ is the center-to-center separation between neighboring nanomagnets, $\theta_n$ is the angle subtended by the $n$-th nanomagnet's magnetization vector with the x-axis, $K_1$ is the first order magnetocrystalline anisotropy constant, and $\lambda_{100}$ is the magnetostriction constant.

The material parameters for nickel are given in Table I. We adopt the convention that tensile stress is positive and compressive stress is negative. The PZT layer can transfer up to a strain of 500 ppm to the Ni layer [27], so that the maximum stress that can be generated in that layer is 100 MPa. The shape of the nanomagnets is that of a circular disk of diameter of 100 nm and thickness 10 nm, while the center-to-center separation between the nanomagnets is $R = 160$ nm. These dimensions ensure that the nanomagnet is single-domain [28]. The parameters are chosen such that: (i) the magnetocrystalline anisotropy energy barrier is 0.55 eV (or 22 $kT$) at room temperature. This makes the static error probability associated with spontaneous flipping of magnetization very small, (ii) the dipole interaction energy is 0.2 eV, which is nearly 3 times smaller than the magnetocrystalline anisotropy energy, and (iii) the stress anisotropy energy at the maximum stress of 100 MPa is 1.5 eV which is enough to overcome the magnetocrystalline energy barrier and make the nanomagnet switch from one orientation to another.

In order to show that the magnetizations of nanomagnets 2 and 3 indeed rotate when the input nanomagnet is switched and the stress cycle on nanomagnets 2 and 3 is executed, and to find the new orientations of these nanomagnets, we follow the procedure below. For each value of stress, find $E_{total-2}$ for every $\theta_2$ while holding $\theta_1$ and $\theta_3$ constant at their initial values. Next, find $E_{total-3}$ versus $\theta_3$ while holding $\theta_4$ constant at the intial value and $\theta_2$ constant at the value corresponding to the minimum of $E_{total-2}$. Next, we re-evaluate $E_{total-2}$ versus $\theta_2$ while changing $\theta_3$ to the value corresponding to the minimum of $E_{total-3}$. This process is iterated until convergence is reached.

We now consider the arrangement in row I of Fig. 4, where no stress is applied initially. This is an anti-ferromagnetic arrangement with the input nanomagnet 1 in the "up" state. Accordingly, the initial conditions are $\theta_1 = +90°$, $\theta_2 = -90°$, $\theta_3 = +90°$ and $\theta_4 = -90°$. When the input is flipped, from "up" to "down" ($\theta_1 = -90°$), nanomagnet 2 finds itself in a tie-state (frustrated) since it experiences equal and opposite dipole magnetic fields from magnet 1 and nanomagnet 3. This can be seen in the energy profile of nanomagnet 2 in Fig. 2(a) (the bottom curve) before stress is applied. The profile is symmetric about $\theta_2 = 0°$; hence $\theta_2 = \pm 90°$ are degenerate in energy. In other words, magnet 2 cannot lower its energy by responding to the input, so that it does not respond. At this point, the clocking cycle is initiated to break the tie. The energy profiles of nanomagnets 2 and 3 as a function of their orientation are shown in Fig. 2 with increasing or decreasing compression or tension. The stress cycle consists of Tension (Fig. 2(a))

→ Relaxation (Fig. 2(c)) → Compression (Fig. 2(e)) → Relaxation (Fig. 2(g)) on nanomagnet 2, and simultaneously Compression (Fig. 2(b)) → Compression (Fig. 2(d)) → Tension (Fig. 2(f)) → Tension (Fig. 2(h)) on nanomagnet 3. As noted earlier, the stress is applied along the [100] direction (+45°). This can be accomplished by applying a voltage across the piezoelectric layer, which generates the strain in this layer through $d_{31}$ coupling. Most of this strain is transferred to the nickel layer which is much thinner than the piezoelectric layer. Furthermore, to ensure uniaxial stress along the +45° axis, the multiferroic nanomagnet is mechanically restrained to prevent expansion and contraction along the direction perpendicular to the +45° axis. The two stress sequences (TRCR, CCTT; where T=tension, C=compression, and R=relaxation) are applied on nanomagnets 2 and 3 simultaneously. Stress is increased or decreased in steps of 0.1 MPa. The '*' markers indicate the magnetization orientations of nanomagnets 2 and 3 in their energy minima for any given stress. The squares identify the final orientation into which the nanomagnet settles at the end of the stressing or relaxation cycle, while circles identify initial orientations. The thin (thick) solid curve represents the energy landscape of a nanomagnet at the onset (end) of a stage of the clock cycle while the dotted lines represent the intermediate energy profiles.

In the first stage of the clock, a tensile stress is applied on nanomagnet 2 (Fig. 2(a)) while a compressive stress is applied on nanomagnet 3 (Fig. 2(b)). The magnetization of nanomagnet 2 rotates from its initial -90° orientation as the tensile stress on it is increased and finally settles to ~ -40° at +100 MPa stress; nanomagnet 3 rotates from +90° to ~ +45° as it is compressed to -100 MPa. It can be seen that at a certain stress (~ 50 MPa), both nanomagnets are drawn towards 0°. This is due to the dipole coupling between the magnets which prefers ferromagnetic coupling over anti-ferromagnetic coupling. Further increase in the stress (tension in nanomagnet 2, compression in nanomagnet 3) results in the nanomagnets settling in their final states at the end of the stage (100 MPa) because the stress anisotropy energy dominates both the dipole and magnetocrystalline energies.

The next stage of the clock cycle involves relaxing the tensile stress on nanomagnet 2 to zero (Fig. 2(c)) while holding the compressive stress on nanomagnet 3 at -100 MPa (Fig. 2(d)). As the stress anisotropy energy in nanomagnet 2 subsides to zero, the relative influence of the dipole energy (due to interaction with neighboring nanomagnets 1 and 3) increases and causes a magnetization rotation from $\theta_2 = \sim -40°$ to $0°$. This rotation towards $0°$ is preferred over a rotation back to -90° since the orientation at 0° is at a lower energy state. Another way to explain this rotation is by resolving the magnetic field components of the neighboring nanomagnets along the x- and y- axes and recalling the preference for ferromagnetic coupling over anti-ferromagnetic coupling. Since nanomagnet 3 is still compressed at -100 MPa, its magnetization remains at ~ +45°. Therefore, the x-component (~ +1050 A/m) of the magnetic field due to its interaction with nanomagnet 2 is twice that of the y-component (~ -525 A/m). Magnet 1 is at -90° and, so, its interaction with nanomagnet 2 produces a magnetic dipole field along the +y-axis with magnitude ~ +750 A/m. The net dipole field on nanomagnet 2 is +1050 A/m along the +x direction ($\theta_2 = 0°$) and +225 A/m along the +y direction ($\theta_2 = \sim +90°$). This results in the magnetization strongly favoring a rotation to 0°.

In the third stage of the clock, a compressive stress, up to a maximum of -100 MPa, is incrementally applied on nanomagnet 2 (Fig. 2(e)). At the same time, the compressive stress on nanomagnet 3 is relaxed (Fig. 2(f)), following which a tensile stress (up to +100 MPa) is applied. The magnetization of nanomagnet 2 rotates from ~ 0° to ~ +45° since this is the closest energy minimum created by the compressive stress along the +45° direction (the raising of the energy barrier at -45° prevents a rotation to the other energy minimum at -135°). Upon relaxation of the compressive stress on nanomagnet 3, the x-component of the magnetic field it experiences owing to its dipole interaction with nanomagnet 2 exceeds the y-component owing to interaction with nanomagnets 2 and 4. This can be seen in the slight tilt towards 0° in the energy profiles of Fig. 2(f) which results in a magnetization rotation towards 0°. The tensile stress applied subsequently induces a rotation from 0° to ~ -45° as the raising of the energy barrier along +45° prevents the magnetization from rotating to the other energy minimum at +135°.

The final stage consists of relaxing the compressive stress on nanomagnet 2 to zero (Fig. 2(g)), while holding the tensile stress on magnet 3 constant (Fig. 2(h)). Upon examination of the dipole field experienced by nanomagnet 2 owing to its interaction with nanomagnet 1 ($\theta_1 = -90°$) and nanomagnet 3 ($\theta_3 = \sim -45°$), it can be determined that the +y-component of the dipole magnetic field (compelling it to rotate "up" to satisfy anti-ferromagnetic ordering) is greater than the +x-component (forcing it to rotate "right" to assume ferromagnetic ordering). Therefore, the magnetization rotates to the desired "up" or $\theta_2 = +90°$. Note that the energy profiles of nanomagnet 2, when undergoing relaxation in this final stage, appear to show an equal tendency for the magnetization to rotate to either 0° or +90°. This occurs due to the preference for ferromagnetic coupling over anti-ferromagnetic coupling. The +90° orientation is ultimately preferred in this case since the dipole magnetic field that would induce a rotation to the "up" state is stronger, albeit slightly, than that which forces a rotation to the "right".

The clocking scheme described above is then repeated on the next set of nanomagnets (nanomagnets 3 and 4) starting with nanomagnet 3 being held under tensile stress. Successive repetition of the clocking cycle on successive sets propagates the input bit unidirectionally down the chain.

In this example, we have shown that the clocking cycle can indeed propagate bits unidirectionally in one case, which corresponds to the first case in Fig. 3. There are eleven more cases to consider. We have considered each one of them and found that the same stress cycle works for all of them.

## 4. CONCLUSION

In conclusion, we have demonstrated an effective clocking scheme that propagates the magnetization state (two logic bits) of a four-state multiferroic nanomagnet unidirectionally along a linear chain by applying a sequence of stresses pairwise on succeeding nanomagnets. This makes it possible to implement multistate logic circuits with wiring connections, fan-out and fan-in. This type of logic circuit is attractive not just because of the higher logic density (4-state versus the usual 2-state), but also because the 4-state elements can be used for associative memory and neuromorphic computing.

In this work, we have neglected the effect of thermal fluctuations that can induce switching errors. Those studies will be reported elsewhere.

In our past work, we had shown [13, 14, 16] that a tiny voltage of $V = 200$ mV is sufficient to generate the maximum stress of 100 MPa in the nickel layer, if we choose the PZT layer thickness as 40 nm and the nickel layer thickness as 10 nm. The capacitance $C$ of such a structure with circular cross-section of 100 nm diameter is ~ 2 fF if we assume that the relative dielectric constant of PZT is 1000. Hence, the energy dissipated in a clock cycle to alternate between no stress to compressive to tensile to no stress is $(1/2)CV^2 + 2CV^2 + (1/2)CV^2 = 3 CV^2 = 0.24$ fJ of energy. Based on previous results [13, 14, 16], we estimate that the switching delay will be less than 1 ns. Hence the clock rate can exceed 1 GHz, even when the energy dissipation is so small. The energy dissipation can be reduced even more – down to ~ $10^2$ kT – by appropriate choice of materials [13]. This makes this scheme a fast and high density logic scheme with extremely low energy dissipation. That, coupled with the fact that nanomagnets have no standby power dissipation unlike transistors, makes it an attractive scheme for computing and signal processing.

**Fig. 1:** (a) Planar nanomagnets with uniaxial shape anisotropy are arranged in a line along the in-plane hard axis: (i) The array has anti-ferromagnetic ordering in the ground state where nearest neighbors have anti-parallel magnetizations; (ii) all magnetizations are forcefully reoriented along the in-plane hard axis by a global magnetic field whose direction is indicated with the thick arrow; (iii) the global field is withdrawn and the orientation of the first nanomagnet is aligned along a chosen direction along the easy axis by an external agent to provide an input bit to the array; (iv) dipole interaction flips the second nanomagnet to assume anti-ferromagnetic ordering, and this effect propagates in a domino-fashion until all magnetizations orient along the easy axes with nearest neighbors having anti-parallel magnetizations. (b) When nanomagnets are arranged in a line along their easy axes, they couple ferromagnetically with nearest neighbors having parallel magnetizations.

**Fig. 2:** A multiferroic nanomagnet consisting of strain-coupled Ni/PZT layers viewed from the top. The shape is circular so that the nanomagnet has no shape anisotropy, but it has biaxial magneto-crystalline anisotropy in the nickel layer which is assumed to be a single crystal. This anisotropy produces four stable (minimum energy) magnetization directions (or "easy axes"): "up" (00), "down" (11), "left" (10) and "right" (01), with their respective bit assignments shown in parentheses. The saddle-shaped curve within the circle represents the energy landscape of the nanomagnet in the unstressed ground state. Stress is applied along the +45° [100] axis.

**Fig. 3:** The twelve distinct scenarios encountered during logic propagation. The "Initial State" column shows the the ground state magnetizations of a four-magnet array or "wire" with nanomagnet 1 acting as the input bit to the array. The "Final State" column shows the expected

state of the wire when the input nanomagnet is switched from its initial state to any of the three other possible states. (a) Anti-ferromagnetic arrangement with input = "up", (b) ferromagnetic arrangement with input = "right", (c) anti-ferromagnetic arrangement with input = "down", and (d) ferromagnetic arrangement with input = "left".

**Fig. 4:** The clock cycle and stress sequences involved in propagating a logic bit unidirectionally are illustrated for the anti-ferromagnetic case when the input bit is switched from its initial "up" (row I) to the "down" state, which results in a tie-condition (row II). To counteract this, a 4-stage "clock" cycle is applied to nanomagnets 2 and 3 (rows III – VI) consisting of tension (T), compression (C) and relaxation (R). The stress sequence applied to nanomagnet 2 is T → R → C → R, while nanomagnet 3 undergoes a C → C → T → T stress cycle. At the end of a single clock cycle (row VI), the magnetization of magnet 2 is rotated from its initial "down" state (rows I and II) to the "up" state. The same clock cycle is then repeated on the next set of nanomagnets in the array (3 and 4) to propagate the logic bit further down.

**Fig. 5:** Energy plots of nanomagnets 2 and 3, as a function of the magnetization angles $\theta_2$ and $\theta_3$. The array is initially in the anti-ferromagnetic configuration and the input nanomagnet 1 is flipped from "up" to "down". Nanomagnets 2 and 3 are then clocked with the stress cycles (TRCR on nanomagnet 2 and CCTT on nanomagnet 3, simultaneously). The '*' markers indicate the magnetization orientation at a particular stress while the square indicates the orientation at the end of a clocking stage. The starting magnetization angles are $\theta_1$ = -90°, $\theta_2$ = -90°, $\theta_3$ = +90° and $\theta_4$ = -90°. (a) Initially, with no stress applied, nanomagnet 2 is at an angle $\theta_2$ = -90°. A gradually increasing tensile stress is applied on this nanomagnet, which makes its magnetization

rotate away from the stress axis towards the closest energy minimum which is at $\theta_2 = \sim -40°$. (b) A gradually increasing compressive stress is applied on nanomagnet 3 causing its magnetization to rotate to ~ 45°. (c) The tensile stress on nanomagnet 2 is gradually relaxed to zero, resulting in its magnetization settling at ~ 0° because of the dipole interaction with nanomagnet 1 (pointing "down") and nanomagnet 3 which is held under compressive stress and whose energy landscape is shown in (d). (e) Nanomagnet 2 is now subjected to a compressive stress which causes it to rotate to ~ +45°, while (f) stress on nanomagnet 3 is relaxed, making it rotate towards 0°. This is immediately followed by a tensile stress that swings its magnetization to ~ -45°. (g) In the final stage of the clock cycle, stress on nanomagnet 2 is relaxed, resulting in its magnetization settling to ~ +90°, while (h) holding the tensile stress on nanomagnet 3 constant.

Table 1: Material parameters for crystalline nickel

| Material parameter | Value |
|---|---|
| Saturation magnetization ($M_s$) | $4.84 \times 10^5$ A/m [25] |
| Magnetostriction constant ($\lambda_{100}$) | $-2 \times 10^{-5}$ [26] |
| Magnetocrystalline constant $K_1$ | $-4.5 \times 10^3$ J/m$^3$ |
| Young's modulus ($Y$) | 200 GPa |

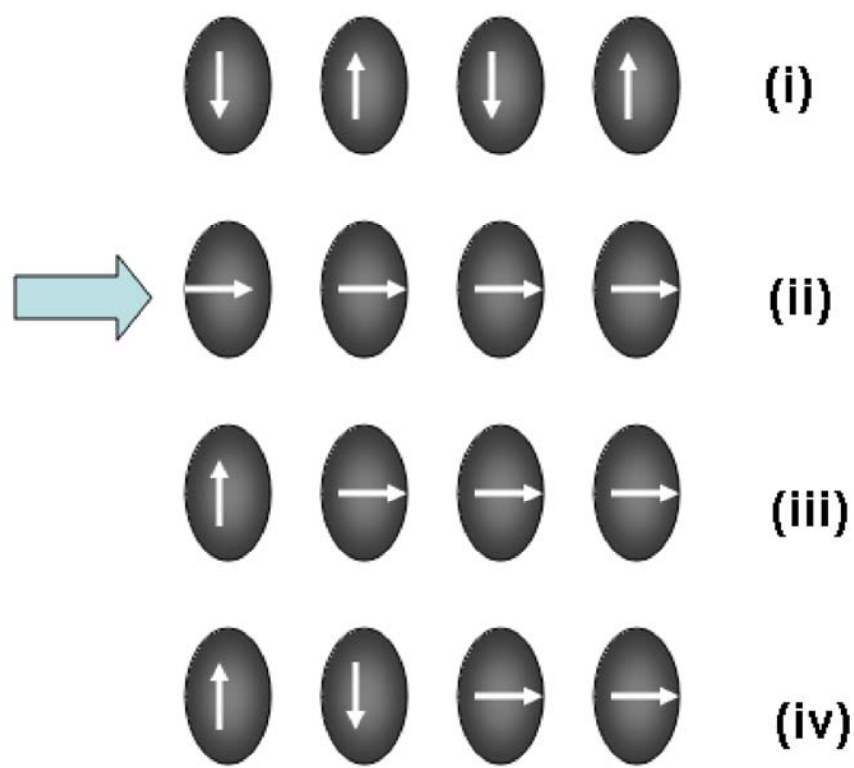

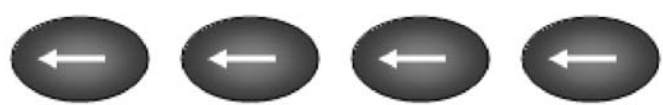

(a)

(b)

**Fig. 1**

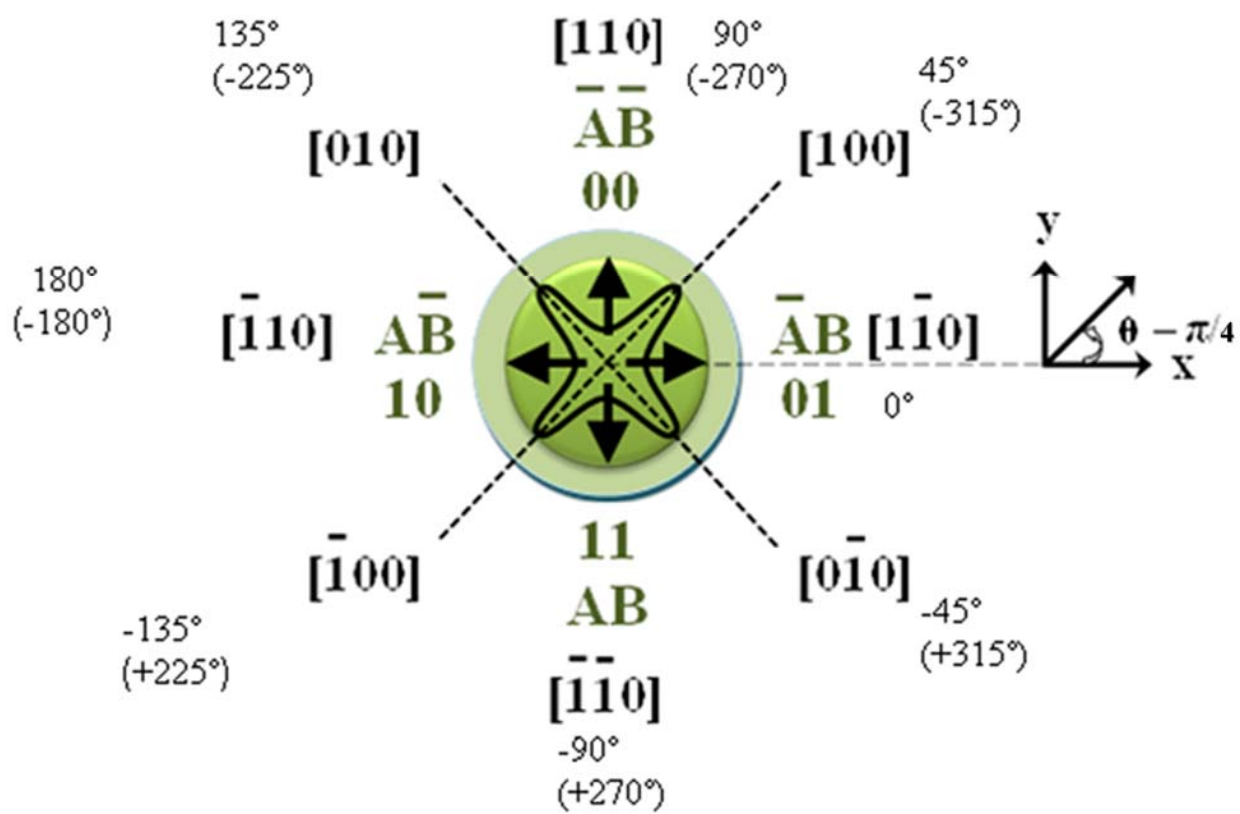

Fig. 2

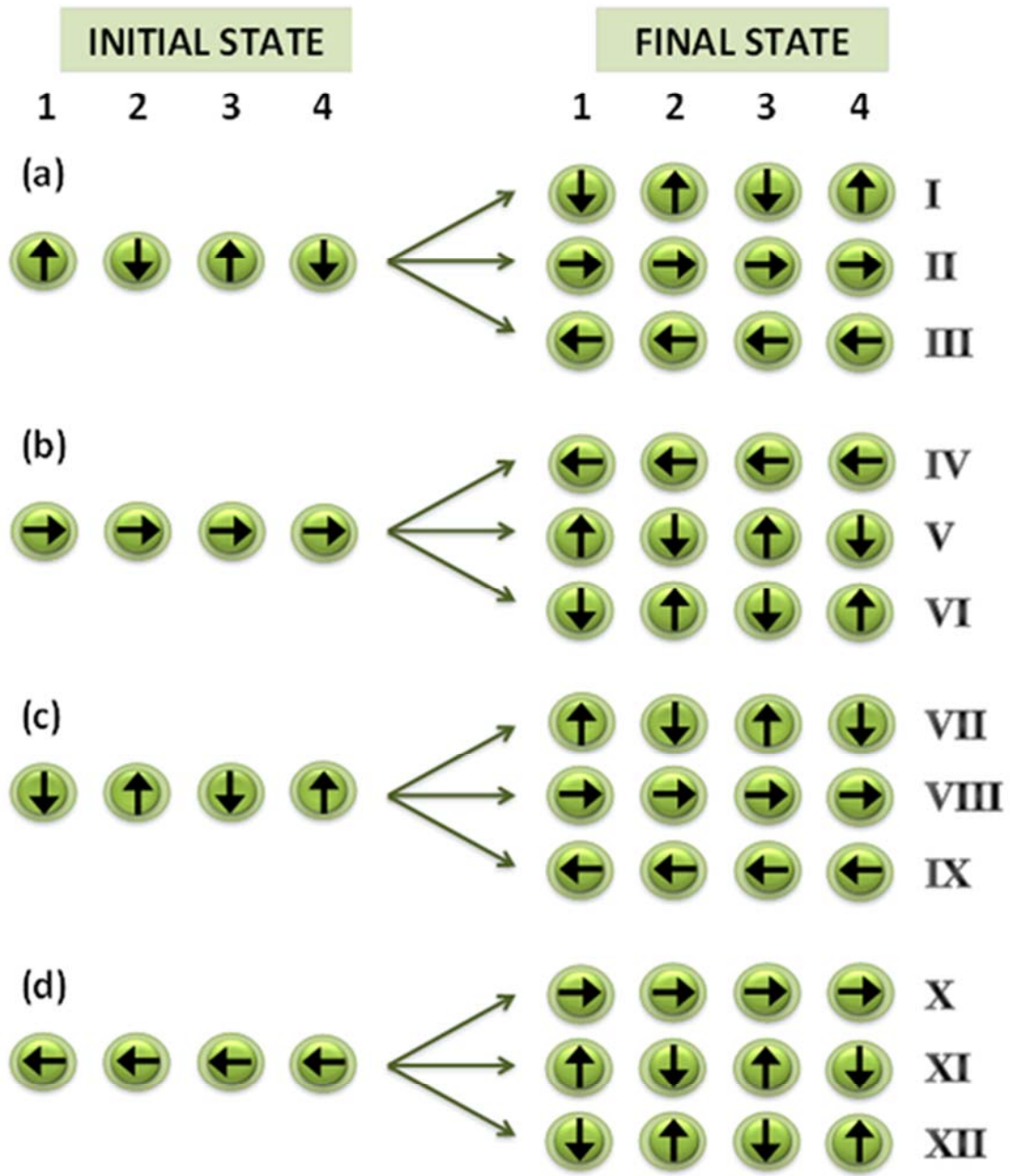

Fig. 3

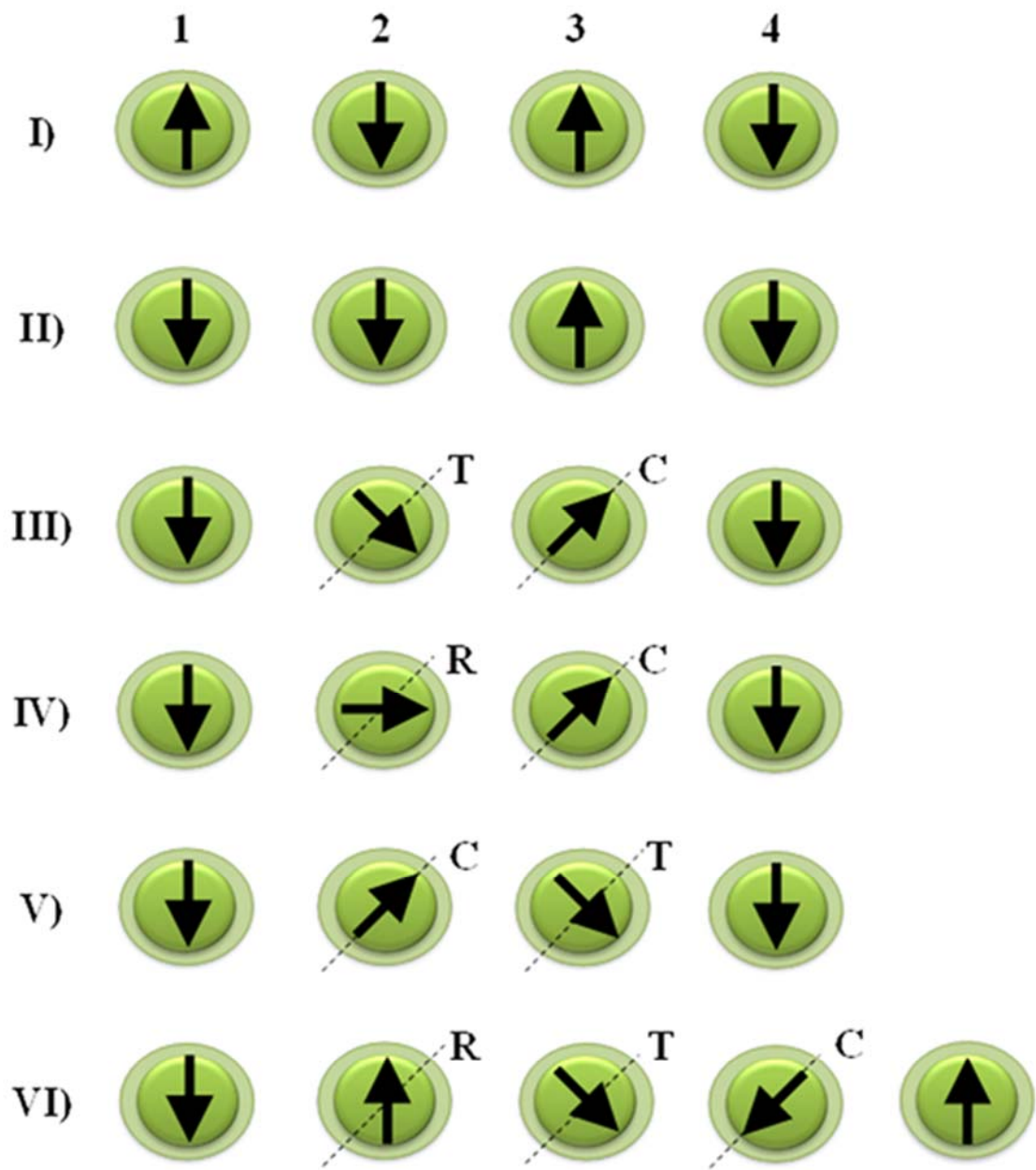

**Fig. 4**

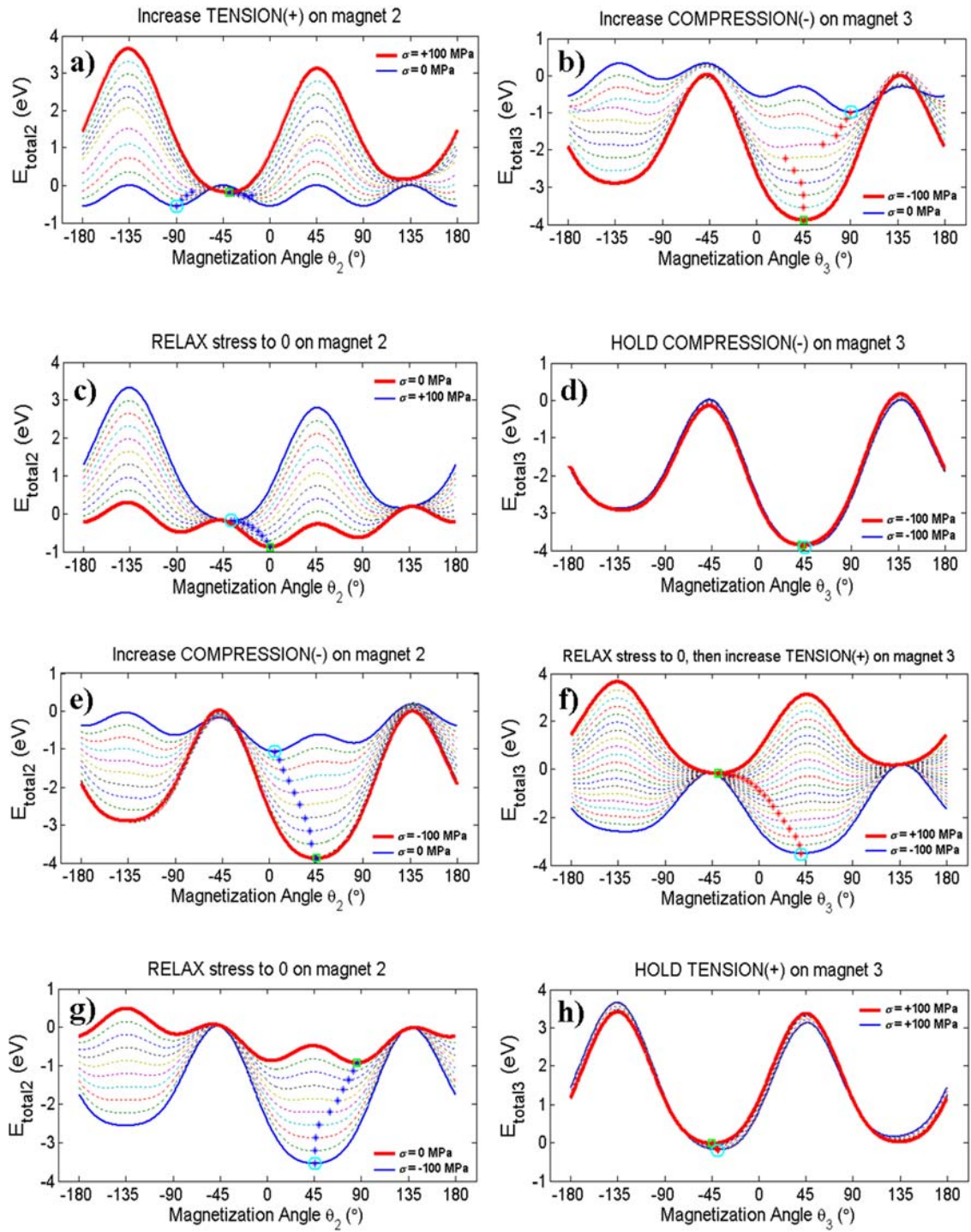

Fig. 5

# SUPPLEMENTARY INFORMATION

# AN ENERGY-EFFICIENT BENNETT CLOCKING SCHEME FOR

# 4-STATE MULTIFERROIC LOGIC

Noel D'Souza[1], Jayasimha Atulasimha[1], Supriyo Bandyopadhyay[2]

1) Department of Mechanical and Nuclear Engineering,

2) Department of Electrical and Computer Engineering,

Virginia Commonwealth University, Richmond, VA 23284, USA.

Email: {dsouzanm, jatulasimha, sbandy}@vcu.edu

In the main paper, we proposed a clocking scheme to propagate four-state nanomagnet logic in Ni/PZT multiferroic nanomagnets having biaxial magnetocrystalline anisotropy. This clocking scheme involves the application of a unique sequence of Compression (C), Tension (T) and Relaxation (R) stresses to the magnetostrictive Ni layer by applying an appropriate electrostatic potential to the PZT layer.

Since each nanomagnet has four stable magnetization orientations, there are twelve possible nanomagnet-array configurations that can arise when the input magnet's orientation is changed from its initial state to any of the three other possible stable states. The possible states of the magnet-array, based on the expectation of ferromagnetic dipole coupling along the array axis and anti-ferromagnetic coupling perpendicular to the array axis, are shown in the 'Final state' column of Fig. 3 in the main paper. Additionally, there are four cases corresponding to the input magnet's magnetization not being changed. In these circumstances, the final magnet states should remain in their original 'ground' states upon completion of the clock cycle. The results included in the main paper described one particular case (Figs. 3(a) – row I), demonstrating the propagation of the magnetization orientation along a nanomagnet array (anti-ferromagnetic arrangement) when the input was flipped from "up" to "down".

We now consider the configuration in which the input magnet is changed from "up" to "right" (Fig. 3(a), row II – main paper). The orientations are $\theta_1 = 0°$, $\theta_2 = -90°$, $\theta_3 = +90°$ and $\theta_4 = -90°$. At this point, the clock cycle we described in the main paper is applied on magnets 2 (TRCR) and 3 (CCTT), whose energy profiles and magnetization rotations are shown in Fig. S1. In the first stage, the tension applied on magnet 2 rotates its magnetization from -90° to ~ -40° (Fig. S1(a)). The compression on magnet 3 rotates its magnetization to +45° (Fig. S1(b)). The strong preference for a rotation towards 0° (higher preference for ferromagnetic coupling than anti-ferromagnetic coupling) can be seen at intermediate stress values (~ 50 MPa). Next, the tensile stress on magnet 2 is relaxed while holding magnet 3 under compression. The magnetic dipole field on magnet 2 due to its interactions with magnets 1 and 3 has a larger +x-component (contributed by magnets 1 and 3) than a –y-component (due to magnet 3 alone) which results in its rotating to 0° (Fig. S1(c)). The next stage of the clock cycle exerts a compressive stress on magnet 2 which rotates its magnetization to from 0° to ~ +40° (Fig. S1(e)). Simultaneously, the compressive stress on magnet 3 is gradually reduced to zero, which rotates its magnetization from +45° to 0°. This is followed by the immediate application of a tensile stress to rotate the magnetization to -45° (Fig. S1(f)). In the last stage of the clock cycle, the stress on magnet 2 is relaxed to zero (Fig. S1(g)) while holding the tensile stress on magnet 3 constant (Fig. S1(h)). Again, the +x-direction is strongly favored and the magnetization of magnet 2 settles into the desired "right" state, reproducing the state of the input bit (magnet 1). By repeating this sequence of stresses on the next set of magnets (3 and 4), the logic (magnetization orientation) is propagated to magnet 3 and beyond when the clock cycle is applied to subsequent magnet pairs.

The next two sets of results correspond to the initial/ground states of the nanomagnet array pointing "right" ($\theta_1 = \theta_2 = \theta_3 = \theta_4 = 0°$) in a ferromagnetic-coupled arrangement.

Fig. S2 shows the energy profiles of magnets 2 and 3 when subjected to the clock cycle following a change in the input magnetization (magnet 1) from "right" to "down". The same stress cycle (TRCR on magnet 2, CCTT on magnet 3) achieves the desired magnetization rotation. In the first stage, magnet 2 rotates from 0° to ~ -40° as a tensile stress of up to +100 MPa is applied to it (Fig. S2(a)). Simultaneously, magnet 3 is gradually compressed to -100 MPa and its magnetization rotates from 0° to ~ +40° (Fig. S2(b)). It can be observed that although the stresses are applied along the +45° direction, the magnetizations of magnets 2 and 3 do not settle at -45° and +45°, respectively. This is because, even at the maximum stress magnitude of 100 MPa, the dipole interaction between nearest neighbors (which prefers a ferromagnetic arrangement) has small but adequate contribution to the total energy of the nanomagnet that biases the orientation slightly away from the -45° and +45° states towards the 0° states. A higher stress will align the magnetization closer to the ±45° axis. When the tensile stress on magnet 2 is relaxed to zero, while keeping magnet 3 compressed, its magnetization settles to ~ 0° (Fig. S2(c)). As explained earlier, this occurs since the *x*-component of the magnetic dipole field of magnet 2 (favoring a parallel alignment with the +*x*-axis) has a stronger contribution to the dipole energy term than the *y*-component (favoring an anti-parallel alignment along the +*y*-axis). Next, magnet 2 is compressed which rotates its magnetization from ~ 0° to ~ +45° (Fig. S2(e)). At the same time, the stress on magnet 3 is relaxed, which causes its magnetization to rotate towards 0°, to be immediately followed by a tensile stress that rotates it to ~ -40° (Fig. S2(f)). Finally, while tensile stress is held on magnet 3 and magnet 2 is relaxed, its magnetization settles to the desired orientation of ~ +90° (Fig. S2(g)). It is driven to this "up" state due to the +y-component of the dipole magnetic field contributed by magnets 1 and 3. Magnet 3 also adds a +x-component that makes the magnetization of magnet 2 want to rotate towards 0°, but its magnitude is smaller than that of the +y-component, which ultimately results in the rotation towards +90°.

The next case considered is when the input magnetization is switched from "right" to "left". Propagation of the input bit through the array due to appropriate magnetization rotations is illustrated in the energy profiles of magnets 2 and 3 in Fig. S3. Prior to initiation of the clock cycle, magnet 2 experiences no net dipole interaction and is in a frustrated state since magnet 1 wants it to flip to the "left" while magnet 3 wants it to stay pointing to the "right". Both influences are equally strong and exactly cancel. This can be seen in the symmetric energy curves of Fig. S3(a) (solid dark blue curve). In contrast, magnet 3 experiences a strong dipole field towards the "right" (towards 0°). When the clock cycle is initiated, a tensile stress is applied to magnet 2 that rotates its magnetization from 0° to ~ -45°, while a compressive stress on magnet 3 causes a rotation from 0° to ~ +45°. In the second stage, relaxing magnet 2 while holding the compressive stress on magnet 3 constant results in a rotation towards -90° (Fig. S3(c)). Next, magnet 2 is compressed to take its magnetization from ~ -90° to ~ -135° (Fig. S3(e)). At the same time, the compressive stress on magnet 3 is relaxed (causing a rotation towards 0°), followed by a tensile stress that rotates its magnetization to ~ -45° (Fig. S3(f)). Finally, magnet 2 is relaxed and its magnetization rotates from ~ -135° to ~ 180°, while the tensile stress on magnet 3 is held at +100 MPa. This is expected since magnet 2 expereinces both a –x-component and a +y-component of the dipole magnetic field that causes a rotation that settles at ~ 180°. The tension held on magnet 3 in the final stage also serves as a transition to the first stage of the next clock cycle (tension on magnet 3, compression on magnet 4). Repeated application of these stress sequences in the clock cycle to subsequent magnet pairs propagates the magnetization state of the input magnet along the array.

The magnetization rotations occurring in the additional configurations shown in Fig. 3 (main paper) are also studied (Figs. S4 – S11) to confirm proper propagation of logic following switching of the input magnetization. It is also essential that the clock cycle does not cause spurious rotations when the input magnetization is not changed. These cases have also been investigated and are illustrated in this supplement (Figs. S12 – S15), through which we confirm that the magnetizations remain unchanged and in the initial 'ground' states at the end of the clock cycle, if the input magnet is unchanged.

For any arbitrary initial arrangement of the magnetizations, whenever the first magnet is switched, the clock cycle is initiated on the next two magnets (T→R→C→R on magnet 2, C→C→T→T on magnet 3). This is then repeated on succeeding pairs. By continuing this sequence, the input logic bit can be propagated unidirectionally down the entire array.

# Figure Captions

**Fig. S1**: Energy profiles of magnets 2 and 3 when subjected to the stress sequences for the case in which for the magnets are in an anti-ferromagnetic configuration with input magnet 1 initially "up" and flipped from "up" to "right". The initial magnetization orientations are: $\theta_1 = 0°$, $\theta_2 = -90°$, $\theta_3 = +90°$ and $\theta_4 = -90°$. (a) Applying a gradually increasing tensile stress on magnet 2 rotates its magnetization from -90° to ~ -40°. (b) Simultaneously, a compressive stress on magnet 3 causes its magnetization to rotate from +90° to ~ +45°. (c) Relaxing the stress on magnet 2 results in its magnetization settling to ~ 0° owing to dipole interactions with magnet 1 ($\theta_1 = 0°$) and (d) magnet 3 which is held under compression. (e) Next, an increasing compressive stress is applied to magnet 2 making it rotate to ~ 40°, while (f) magnet 3 is relaxed, inducing a rotation towards 0°. This is immediately followed by a tensile stress that rotates its magnetization to ~ -40°. (g) The final stage involves relaxing the stress on magnet 2 that results in its settling at the desired orientation of ~ 0° while (h) the tensile stress is held on magnet 3, setting it up for the next clock cycle which would be applied to magnets 3 and 4.

**Fig. S2**: Energy profiles of magnets 2 and 3 when subjected to the stress sequences for the case in which the magnets have a ferromagnetic initial state (pointing "right") and the input magnetization is switched from "right" to "down". (a) A tensile stress is applied to magnet 2 that causes its magnetization to rotate from 0° to ~ -40°. (b) Magnet 3 is compressed, resulting in a magnetization rotation to ~ +40°. (c) Relaxing the stress on magnet 2 makes its magnetization settle back to ~ 0° while (d) the compressive stress on magnet 3 is held at -100 MPa. (e) In the third stage of the clock cycle, a compressive stress on magnet 2 results in a rotation to ~ +45°. (f) At the same time, the stress on magnet 3 is relaxed, causing its magnetization to rotate towards

0°. This is immediately followed by a tensile stress that swings the magnetization to ~ -40°. (g) The final stage involves gradually relaxing the stress on magnet 2 and the desired outcome is achieved when the magnetization rotates to ~ +90°, while (h) the tensile stress on magnet 3 is held at +100 MPa.

**Fig. S3**: Energy profiles of magnets 2 and 3 as a function of magnetization angle for the ferromagnetic arrangement (initially pointing "right") when the input magnet is flipped from "right" to "left" and the clock cycle is applied. (a) In the first stage, a tensile stress gradually applied on magnet 2 sees its magnetization rotate from 0° to ~ -45°, while (b) a compressive stress on magnet 3 causes a rotation to ~ +45°. (c) Relaxing the stress on magnet 2 results in a rotation towards -90°, while (d) the compressive stress on magnet 3 is held at -100 MPa. (e) A compressive stress on magnet 2 rotates its magnetization to ~ -135°. (f) Simultaneously, magnet 3 is relaxed, causing its magnetization to settle near 0°. This is immediately followed by a tensile stress that rotates its magnetization to ~ -45°. (g) The final stage of the clock cycle involves relaxing the stress on magnet 2 which results in the desired final state of ~ -180°, while (h) magnet 3 is kept under tension that holds its magnetization along -45°.

**Fig. S4:** Energy plots of magnets 2 and 3 when the initial ordering was anti-ferromagnetic and the input nanomagnet was switched from "up" to "left". At the end of the clock cycle, magnet 2 settles to ~ +180° or the "left" direction, ensuring successful unidirectional propagation of the input bit.

**Fig. S5:** Energy plots of magnets 2 and 3 when the initial ordering was ferromagnetic and the input nanomagnet was switched from "right" to "up". At the end of the clock cycle, magnet 2

settles to ~ -90° or the "down" direction, ensuring successful unidirectional propagation of the input bit.

**Fig. S6:** Energy plots of magnets 2 and 3 when the initial ordering was anti-ferromagnetic and the input nanomagnet was switched from "down" to "up". At the end of the clock cycle, magnet 2 settles to ~ -90° or the "down" direction, ensuring successful unidirectional propagation of the input bit.

**Fig. S7:** Energy plots of magnets 2 and 3 when the initial ordering was anti-ferromagnetic and the input nanomagnet was switched from "down" to "right". At the end of the clock cycle, magnet 2 settles to ~ 0° or the "right" direction, ensuring successful unidirectional propagation of the input bit.

**Fig. S8:** Energy plots of magnets 2 and 3 when the initial ordering was anti-ferromagnetic and the input nanomagnet was switched from "down" to "left". At the end of the clock cycle, magnet 2 settles to ~ +180° or the "left" direction, ensuring successful unidirectional propagation of the input bit.

**Fig. S9:** Energy plots of magnets 2 and 3 when the initial ordering was ferromagnetic and the input nanomagnet was switched from "left" to "right". At the end of the clock cycle, magnet 2 settles to ~0° or the "right" direction, ensuring successful unidirectional propagation of the input bit.

**Fig. S10:** Energy plots of magnets 2 and 3 when the initial ordering was ferromagnetic and the input nanomagnet was switched from "left" to "up". At the end of the clock cycle, magnet 2 settles to ~ -90° or the "down" direction, ensuring successful unidirectional propagation of the input bit.

**Fig. S11:** Energy plots of magnets 2 and 3 when the initial ordering was ferromagnetic and the input nanomagnet was switched from "left" to "down". At the end of the clock cycle, magnet 2 settles to ~ +90° or the "up" direction, ensuring successful unidirectional propagation of the input bit.

**Fig. S12:** Energy plots of magnets 2 and 3 when the initial ordering was anti-ferromagnetic and the input nanomagnet was *not* switched (remains pointing "up"). At the end of the clock cycle, magnet 2 settles to ~ -90° or the "down" direction, ensuring successful unidirectional propagation of the input bit.

**Fig. S13:** Energy plots of magnets 2 and 3 when the initial ordering was ferromagnetic and the input nanomagnet was not switched and remains pointing "right". At the end of the clock cycle, magnet 2 settles to ~ 0° or the "right" direction, ensuring successful unidirectional propagation of the input bit.

**Fig. S14:** Energy plots of magnets 2 and 3 when the initial ordering was anti-ferromagnetic and the input nanomagnet was not switched and remains pointing "down". At the end of the clock

cycle, magnet 2 settles to ~ +90° or the "up" direction, ensuring successful unidirectional propagation of the input bit.

**Fig. S15:** Energy plots of magnets 2 and 3 when the initial ordering was ferromagnetic and the input nanomagnet was not switched and remains pointing "left". At the end of the clock cycle, magnet 2 settles to ~ +180° or the "left" direction, ensuring successful unidirectional propagation of the input bit.

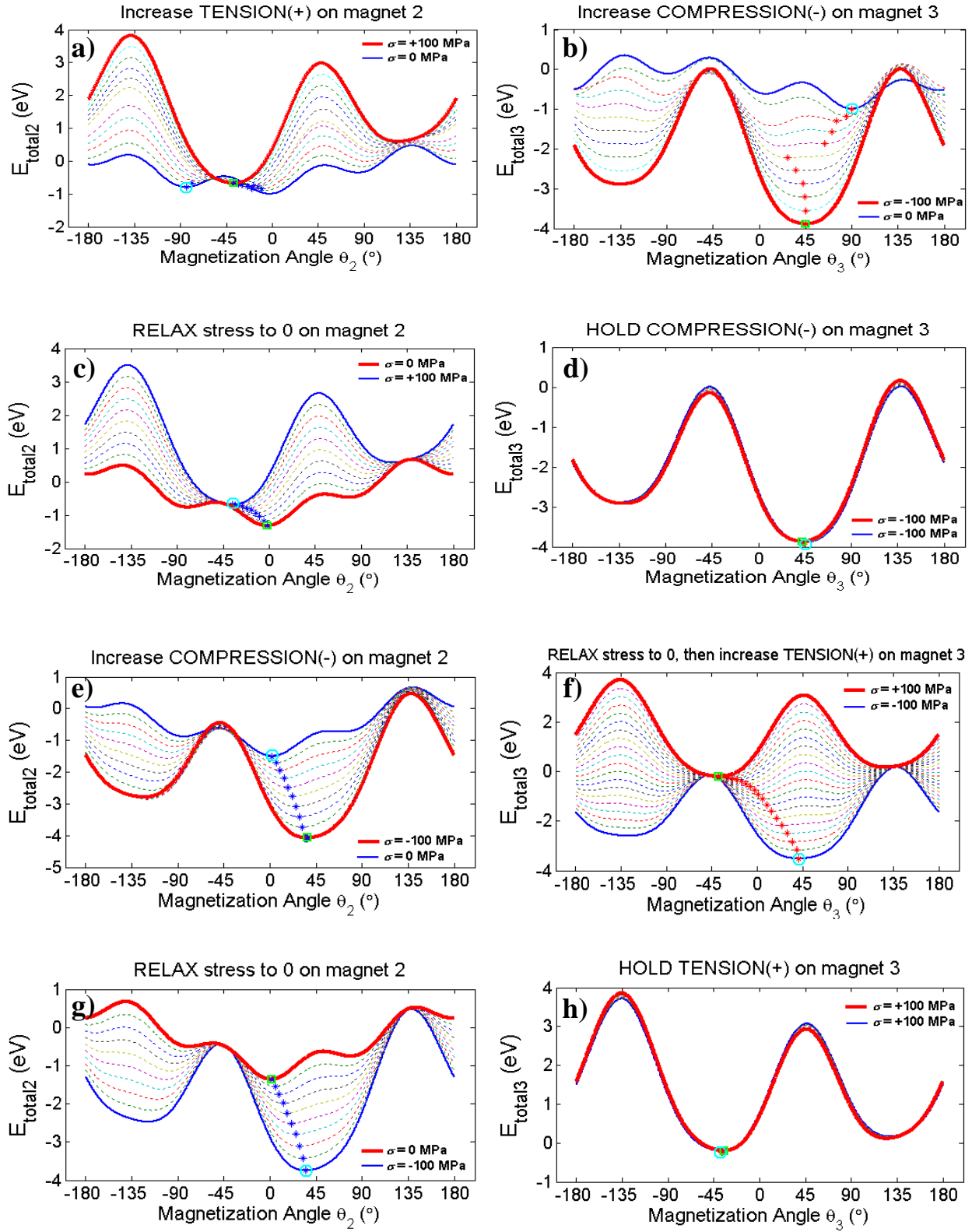

**Fig. S1**

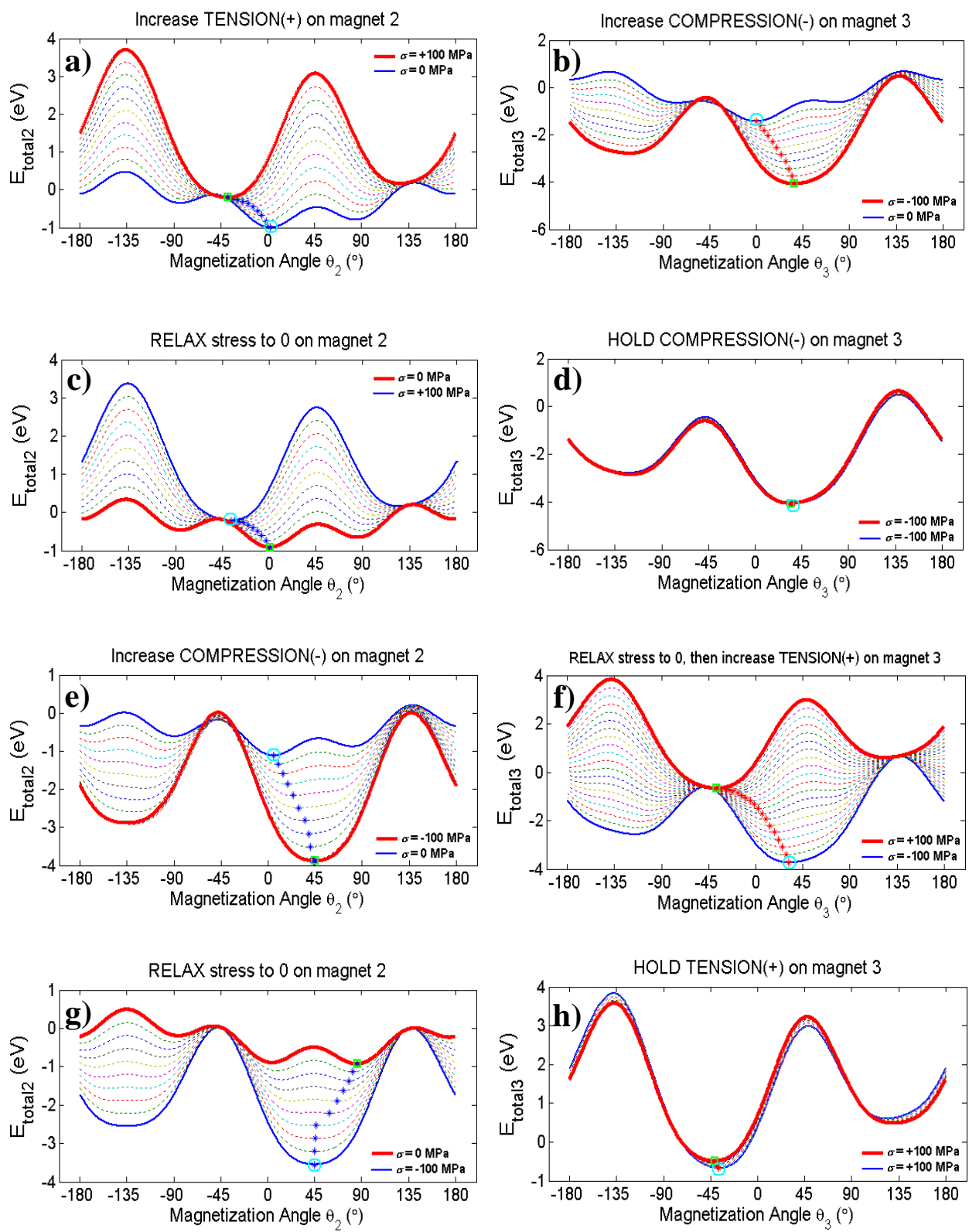

Fig. S2

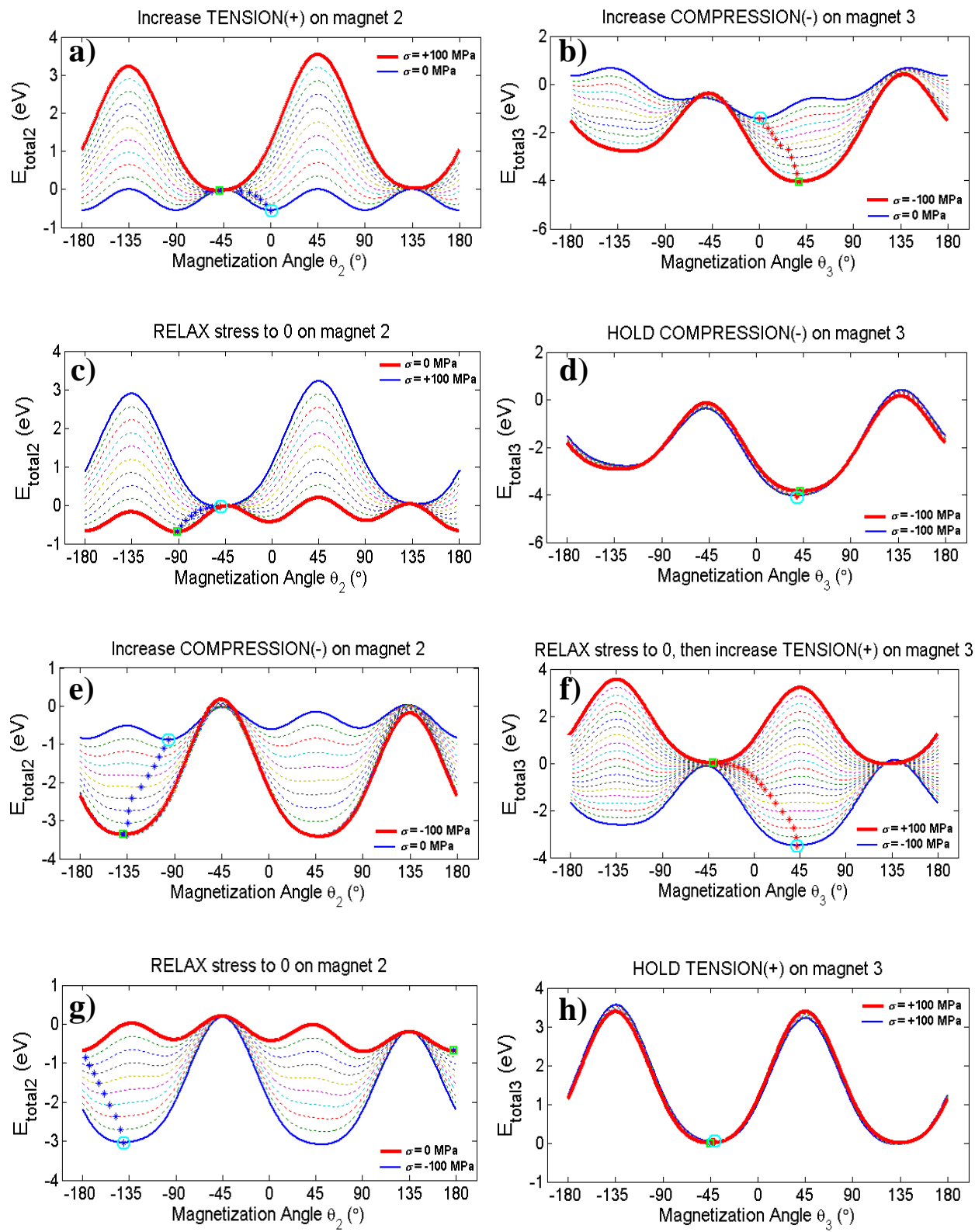

Fig. S3

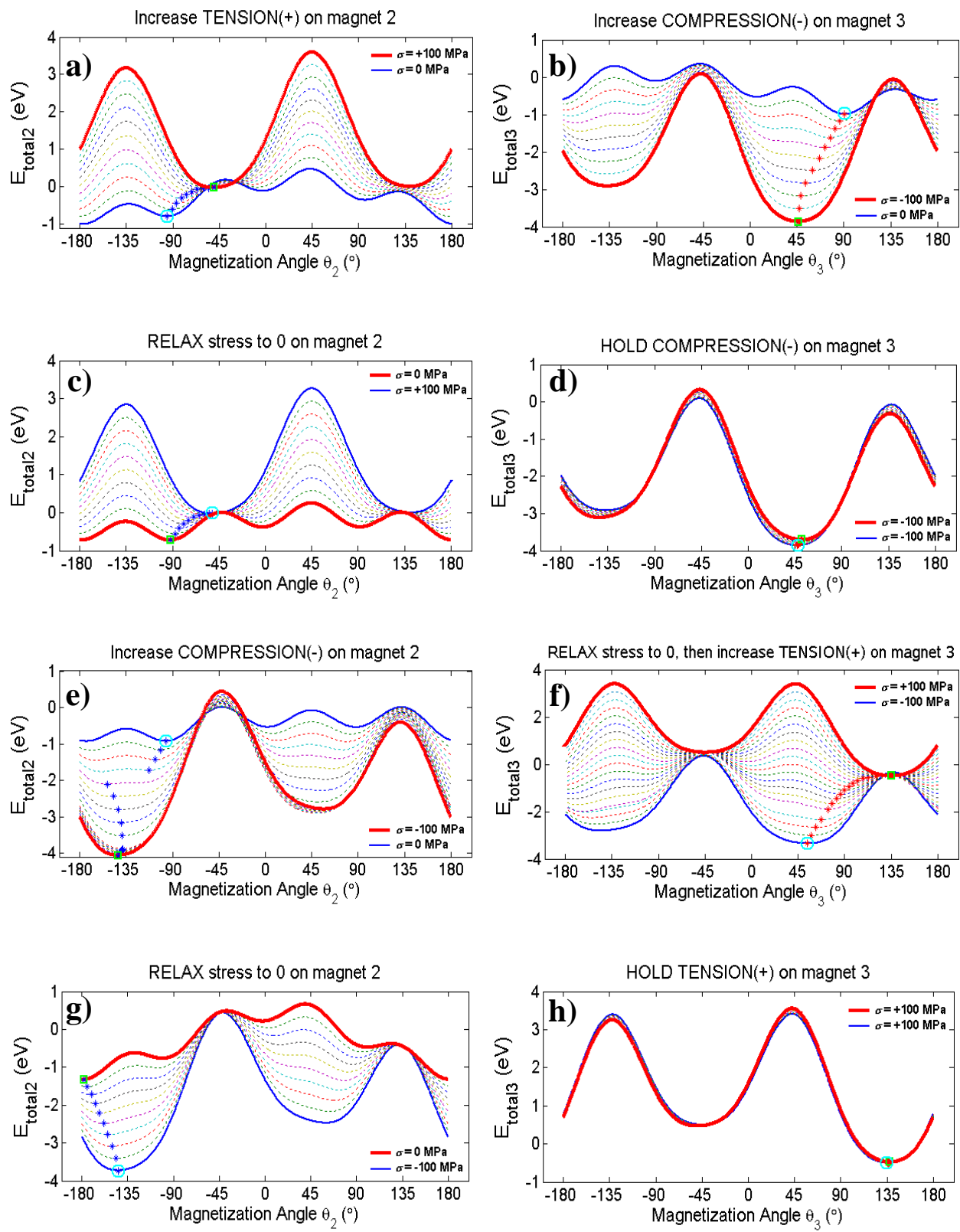

Fig. S4

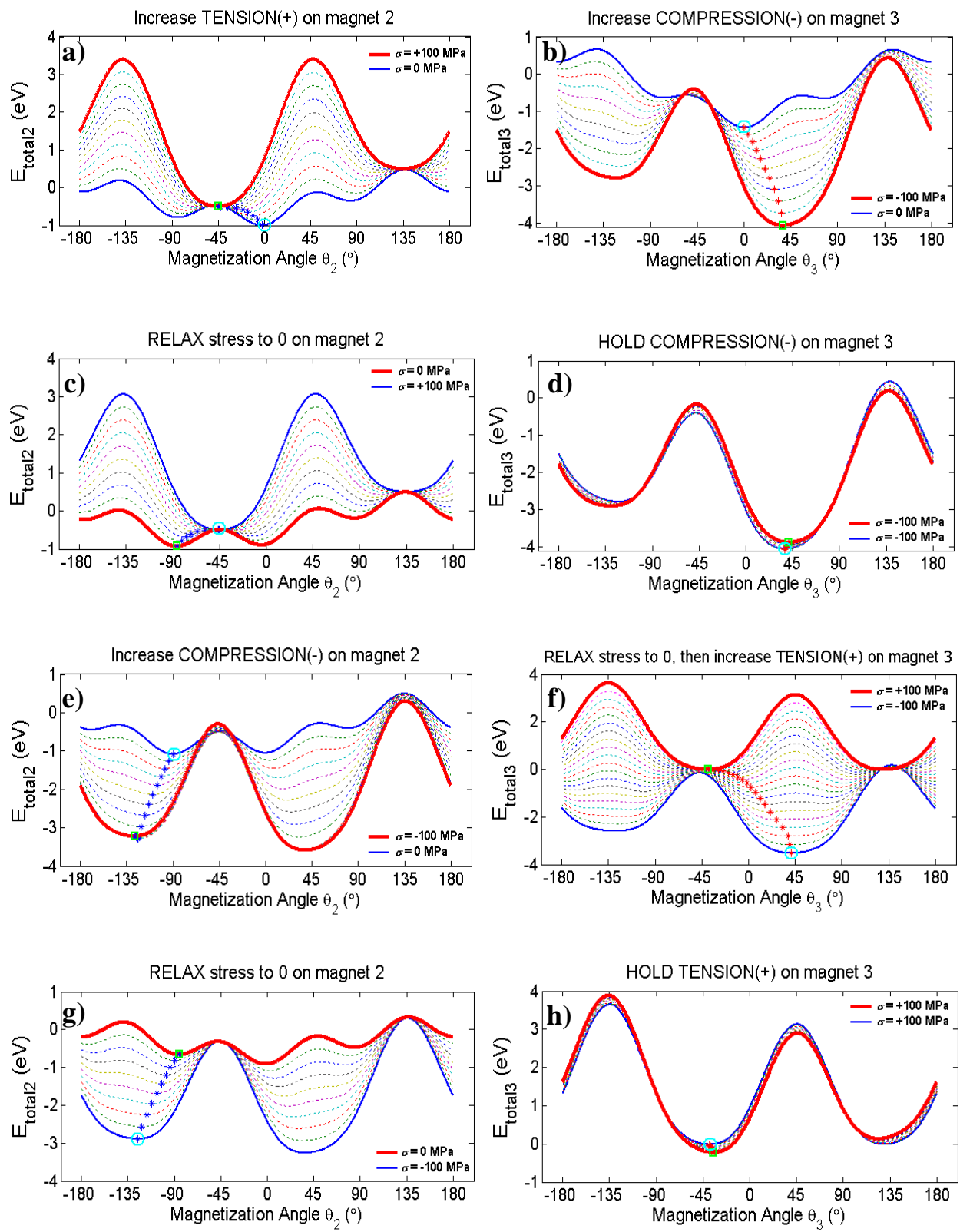

Fig. S5

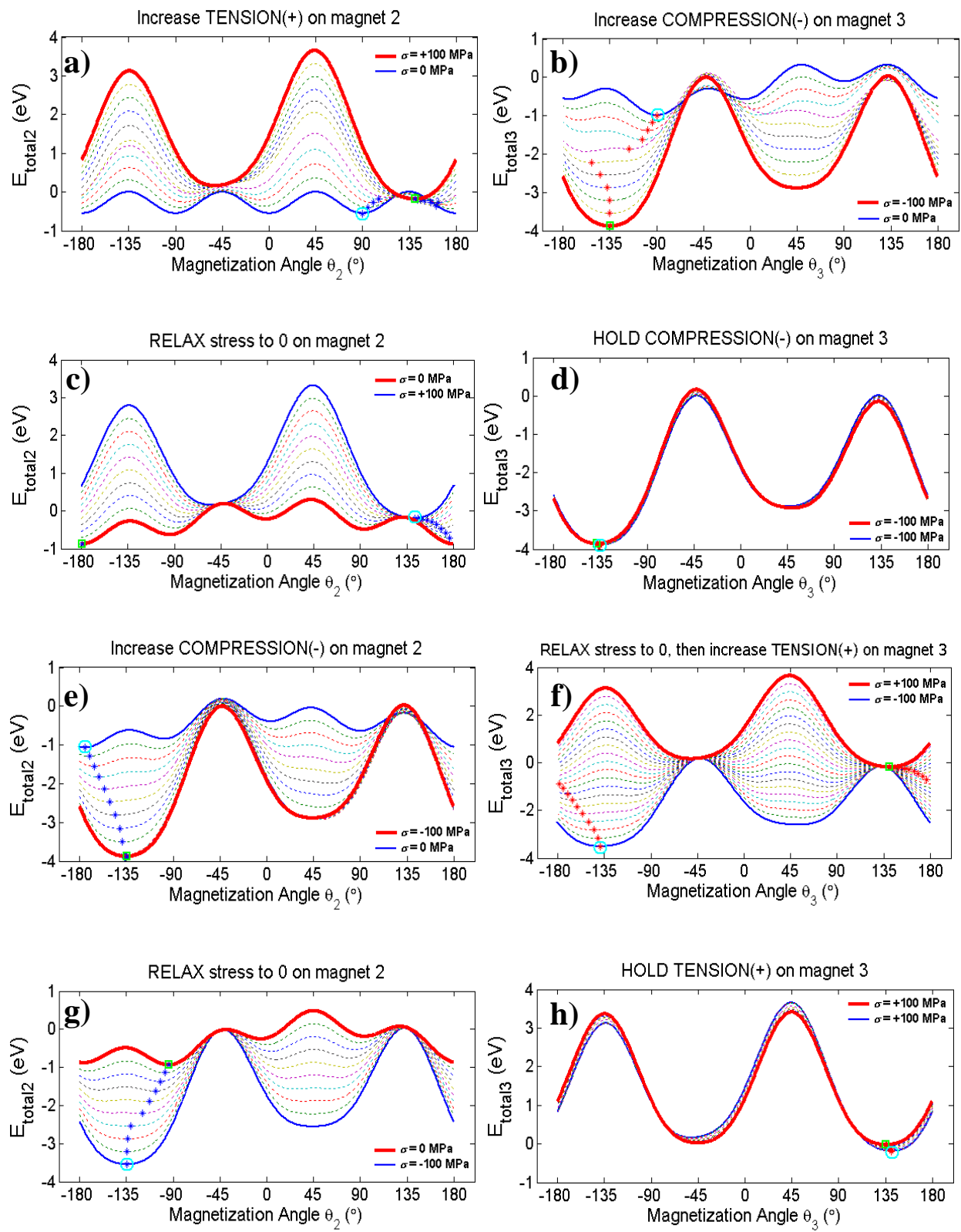

**Fig. S6**

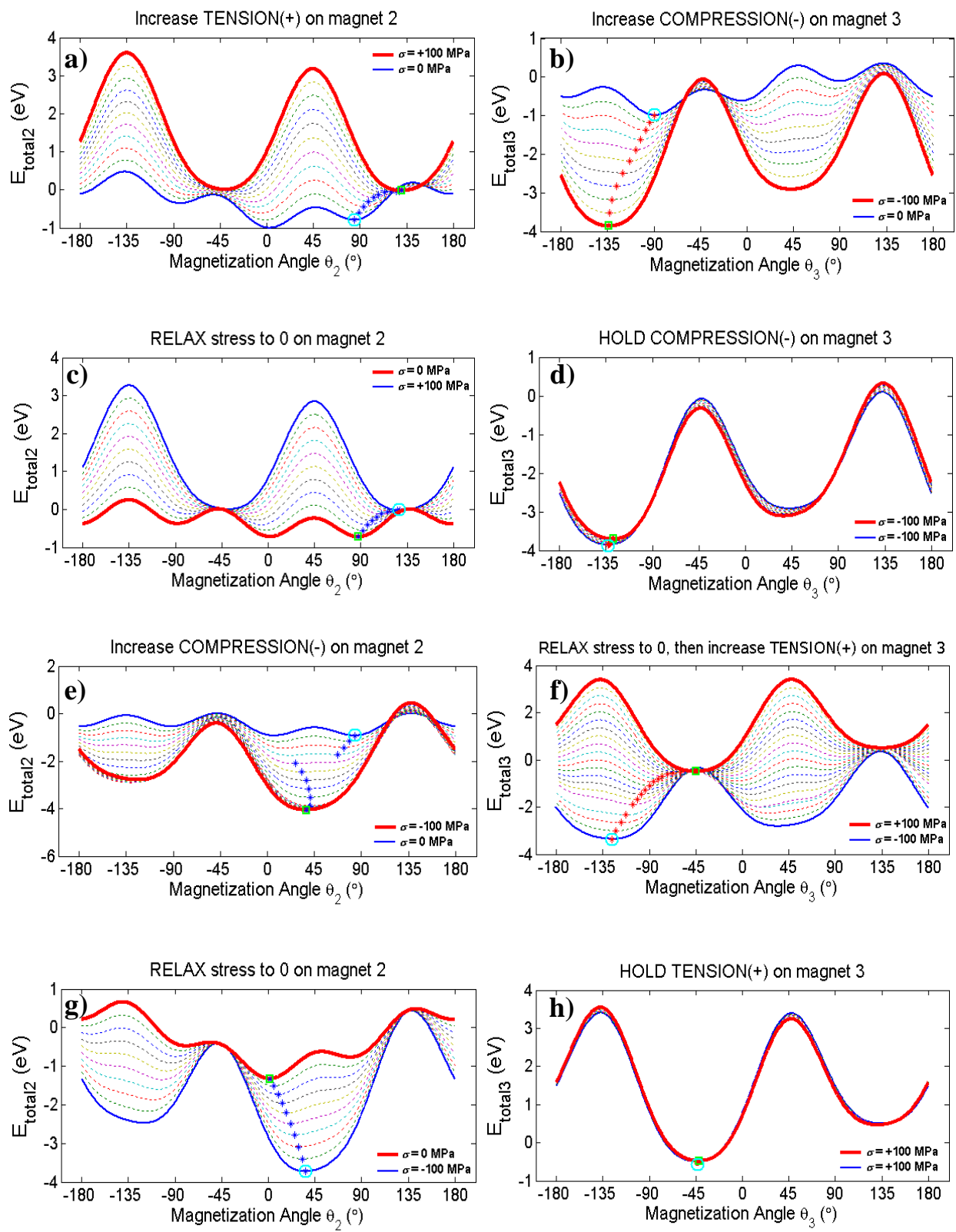

**Fig. S7**

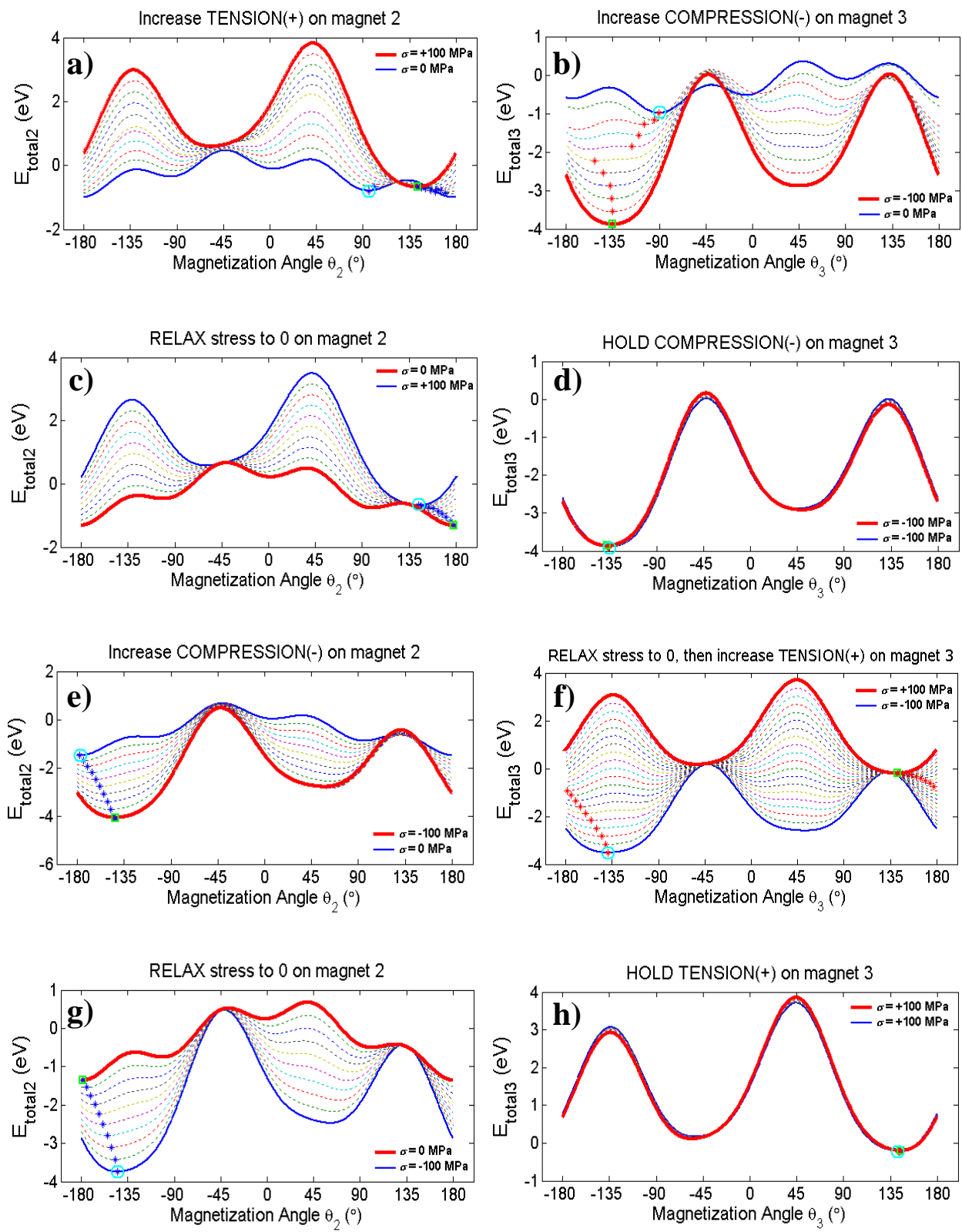

**Fig. S8**

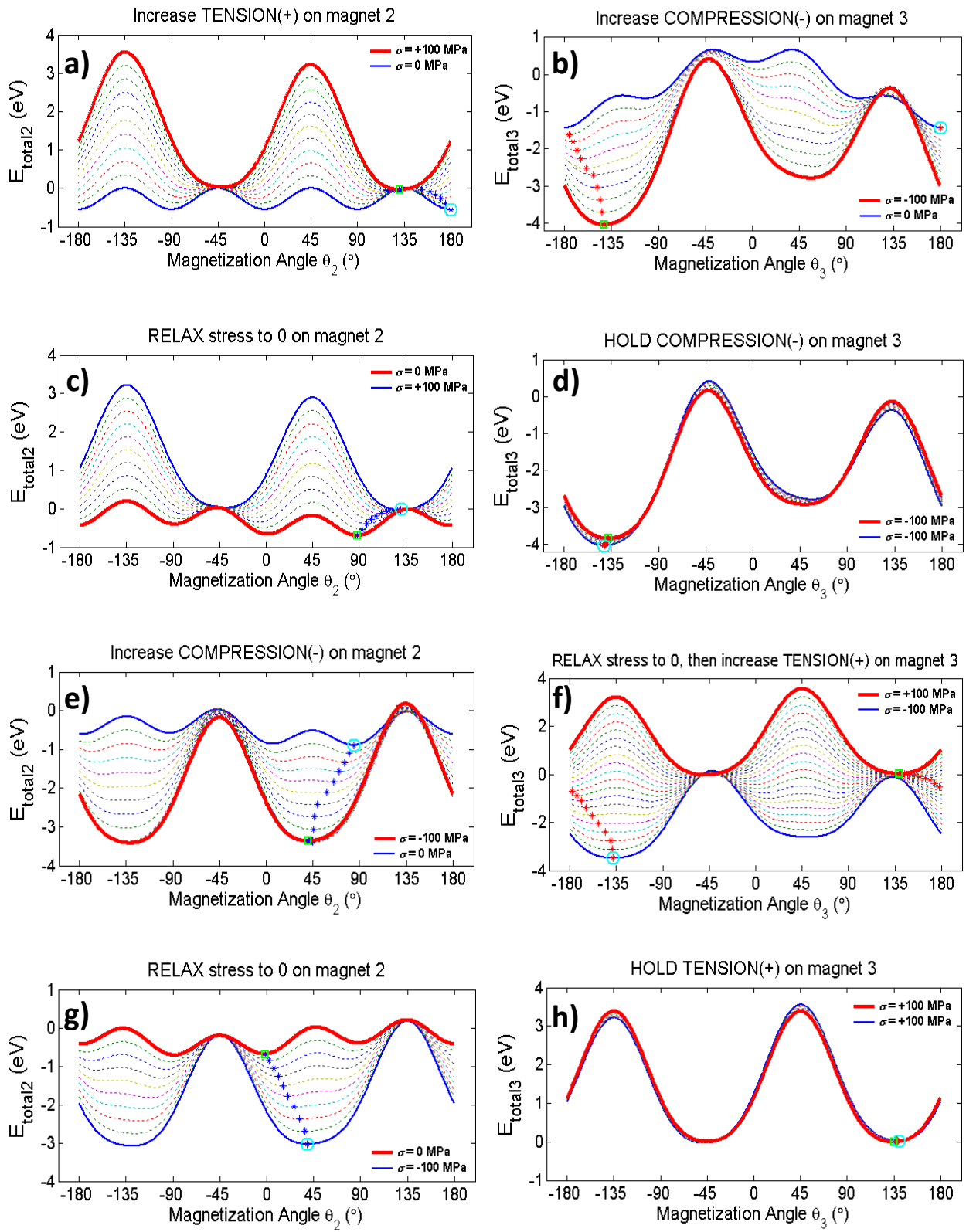

Fig. S9

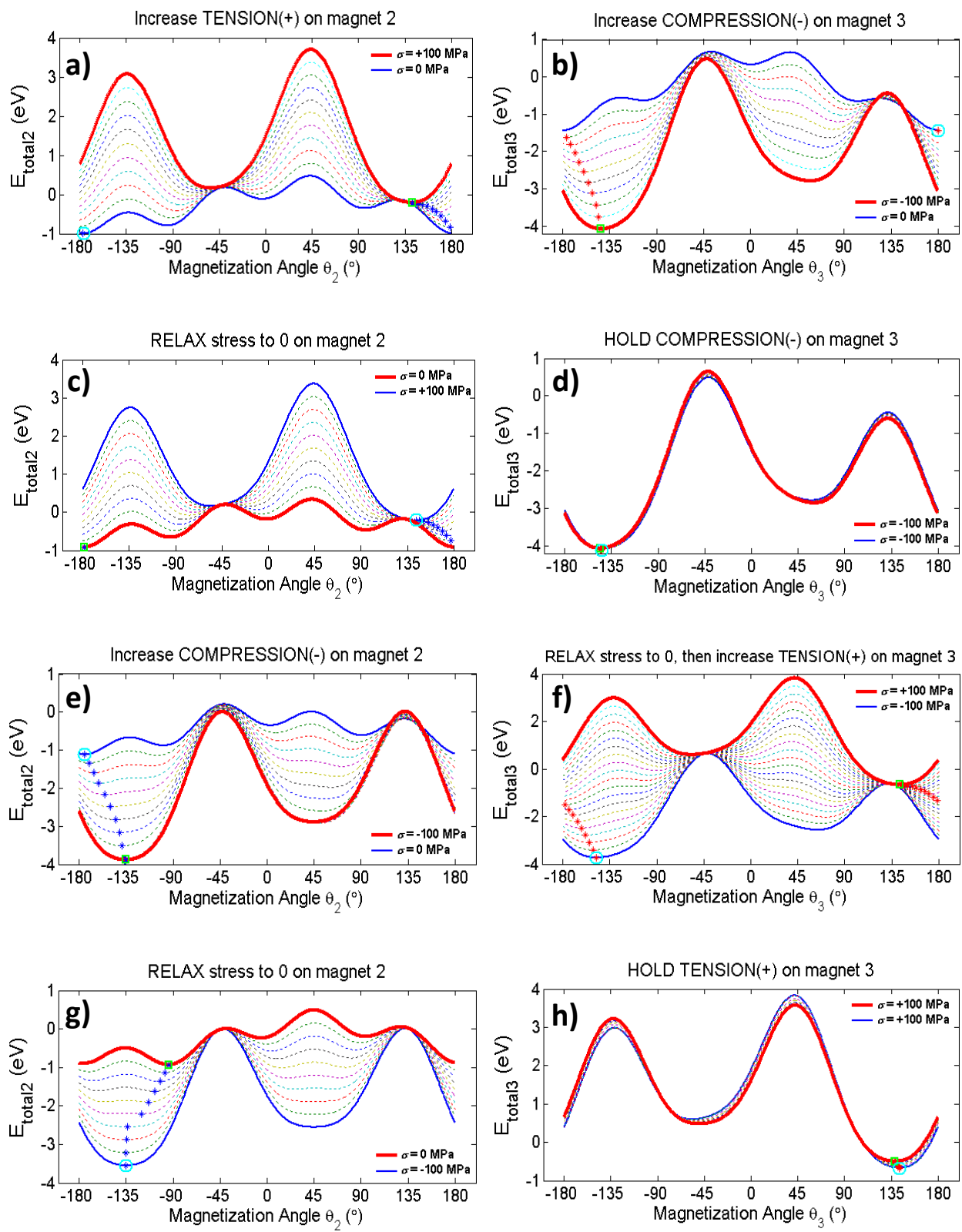

Fig. S10

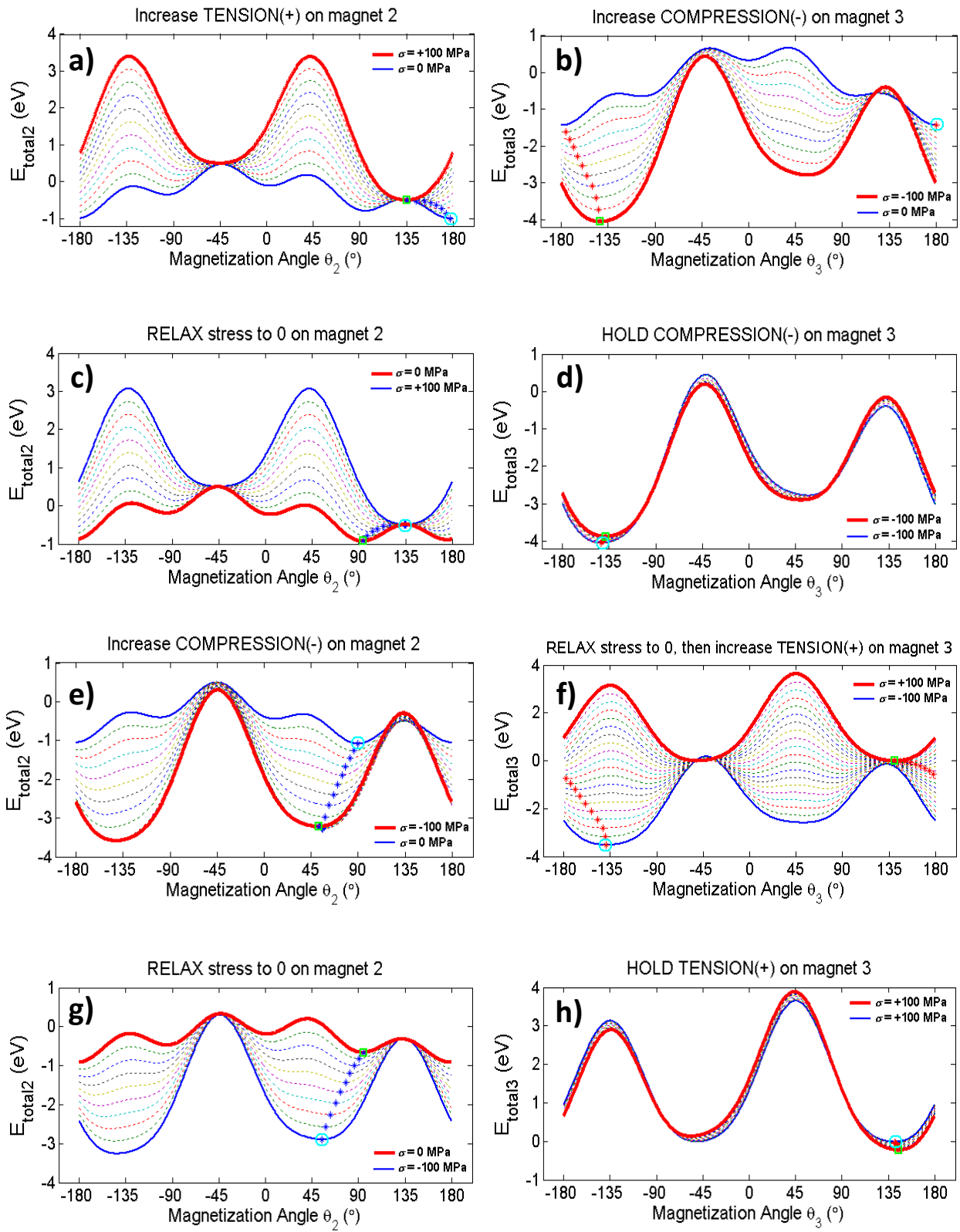

Fig. S11

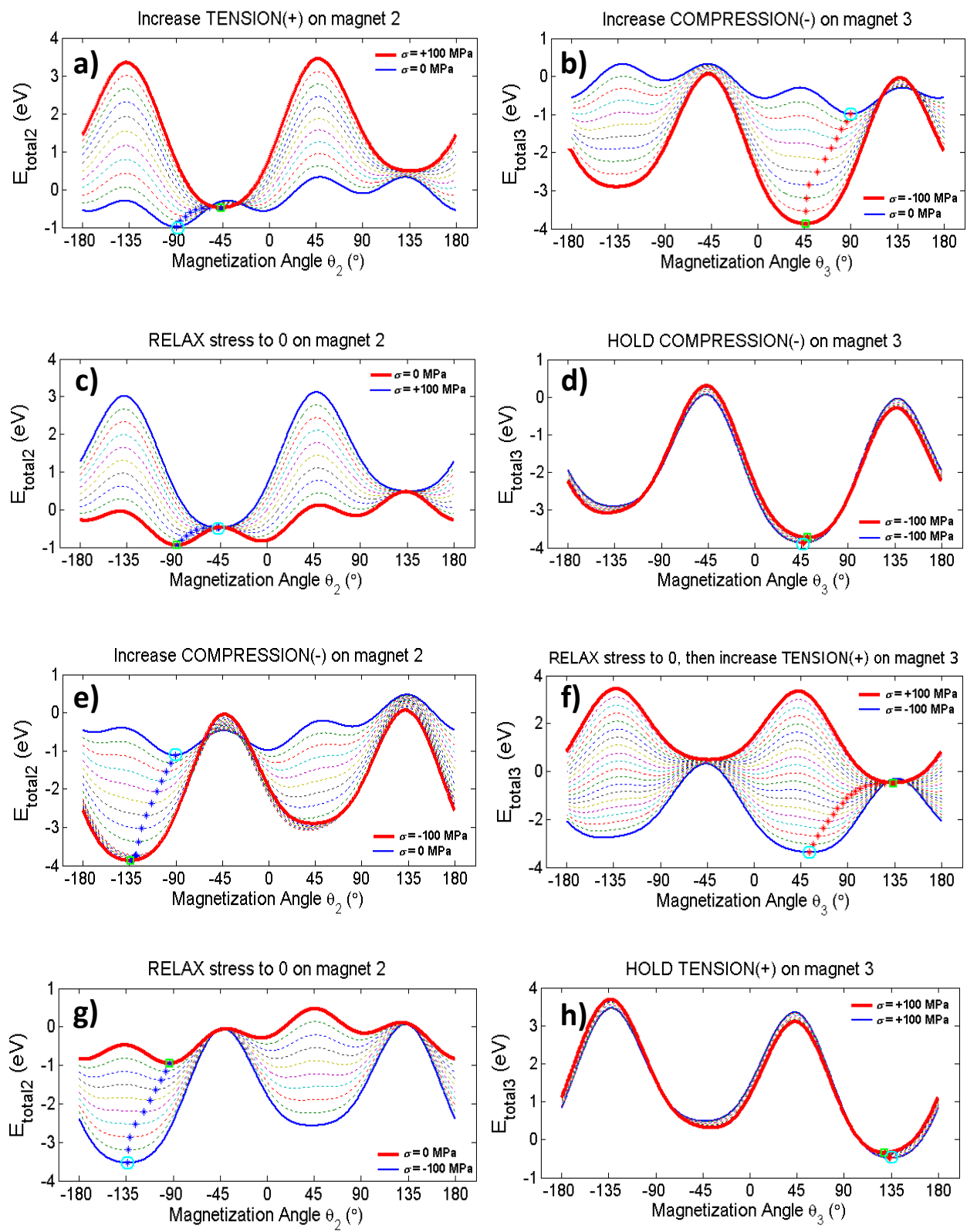

Fig. S12

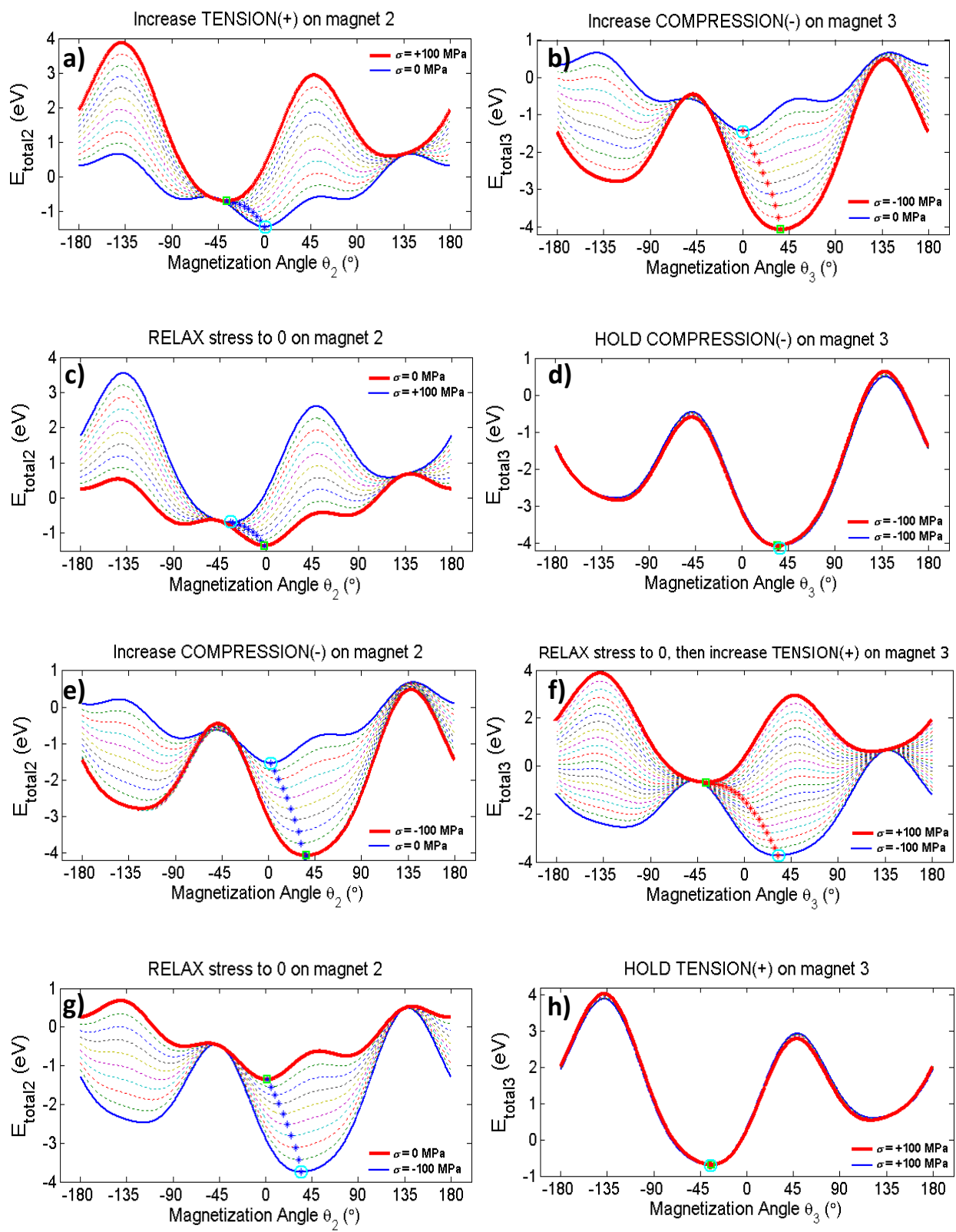

Fig. S13

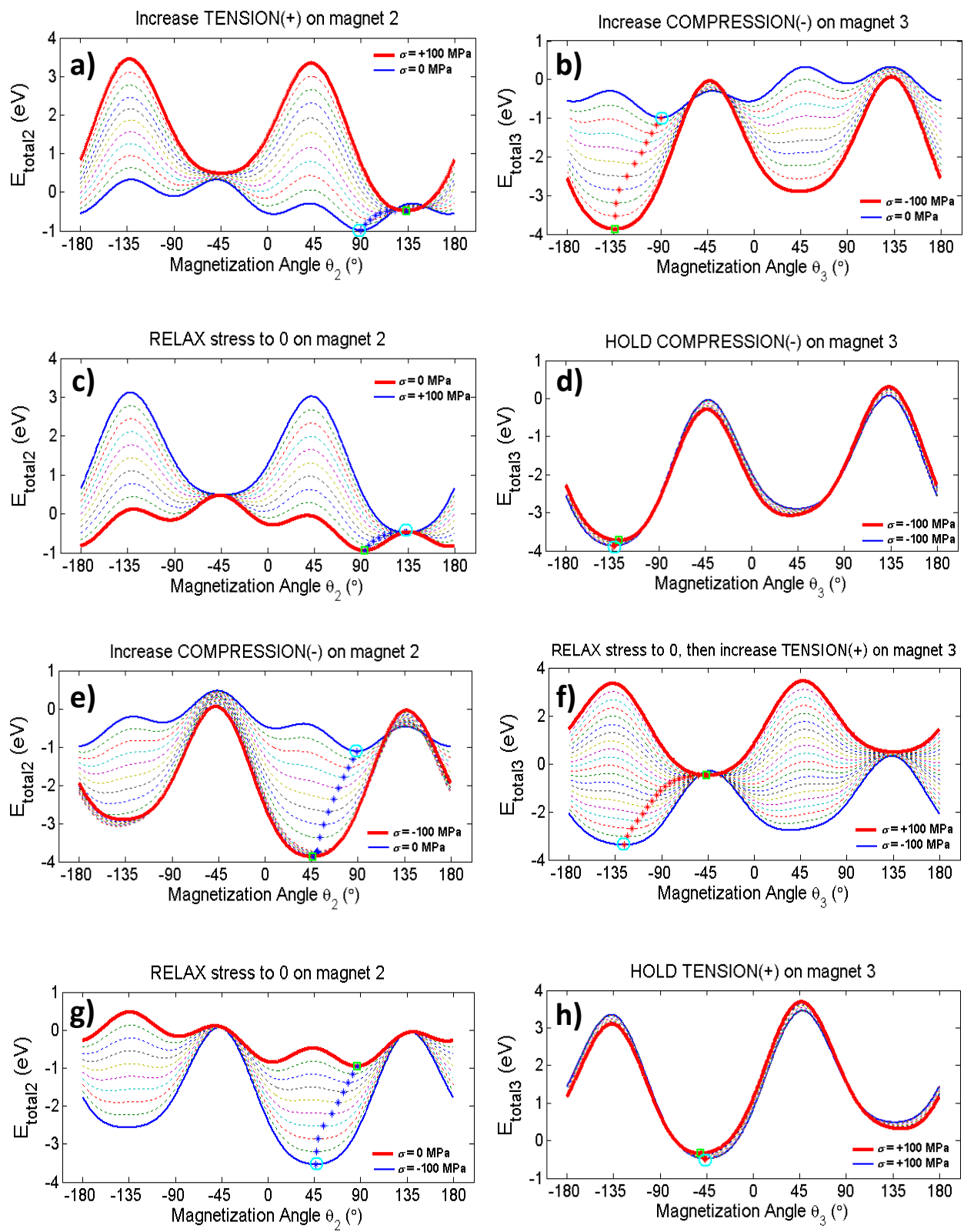

**Fig. S14**

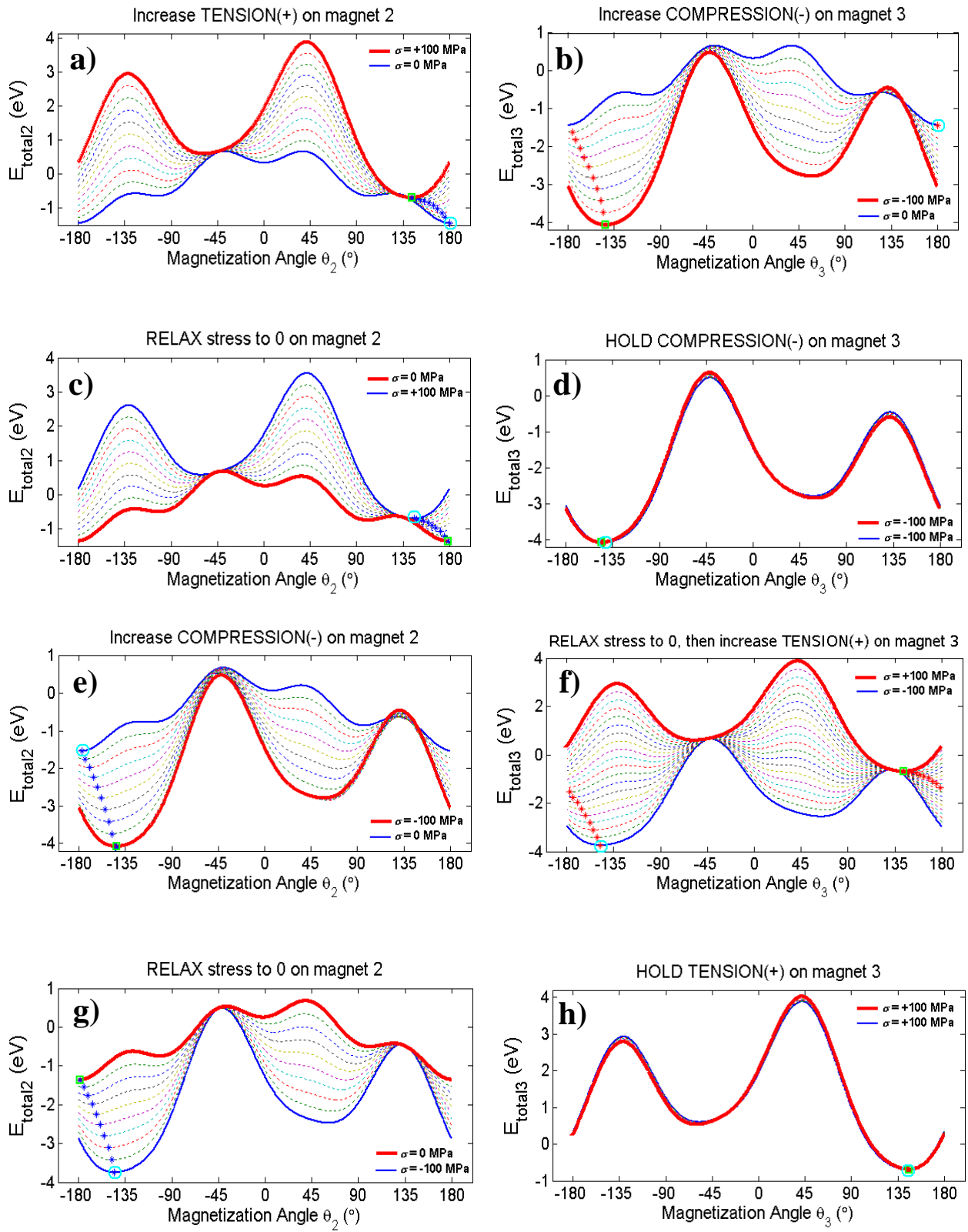

**Fig. S15**